# A Nonlocal Orientation Field Phase-Field Model for Misorientation- and Inclination-Dependent Grain Boundaries

Xiao Han[1] and Axel van de Walle[1]

[1] School of Engineering, Brown University, Providence, Rhode Island 02912, USA

Contact author: xiao_han@brown.edu

Contact author: avdw@brown.edu

**ABSTRACT** We propose to incorporate grain boundary (GB) anisotropy in phase-field modeling by extending the standard partial differential equations (PDE) formulation to include a nonlocal functional of an orientation field. Regardless of the number of grains in the simulation, the model uses a single orientation field and incorporates grain misorientation and inclination information obtained from sampling the orientation field at optimized locations in the vicinity of the grain boundary. The formalism enables simple and precise tuning of GB energy anisotropy while reducing an extensive fitting procedure. The functional includes explicit GB functions to control the GB energy as a function of both misorientation and inclination. The model is validated by reproducing the linear grain growth rate, Wulff shapes with varying misorientations and anisotropic coefficients, and analytical equilibrium dihedral angles at triple junctions. Polycrystalline simulations further demonstrate grain growth, coalescence, triple junction behavior, and the influence of anisotropy on grain morphology.

## I. INTRODUCTION

Phase-field simulations [1-3] are widely used to study interface phenomena in polycrystalline materials. They have proven to be extremely successful in describing microstructure evolution during material synthesis, processing, and service. Yet, producing a realistic phase-field description of polycrystalline materials requires an accurate and efficient description of GB energetics and associated time evolution equations. Raw computing power is no longer the main limiting factor in bridging the gap between atomistic calculations of GB energy and a mesoscopic description of microstructure evolution. The field needs proper theoretical and computational frameworks to make this connection transparently and in full generality.

While considerable progress has been made in incorporating anisotropy in interface energy and mobility, no single scheme has emerged as the preferred method to allow for the completely general anisotropy in misorientation- and inclination-dependence associated with GBs. To understand the challenges faced, it is instructive to overview some of the main existing schemes to incorporate anisotropy.

We first note that inclination-dependence, which would be sufficient to model surface energy or solid-liquid interface, has been previously incorporated into phase-field models [4-10]. One can simply introduce an anisotropy in the direction-dependence of the gradient energy term. Such dependence can be implemented as a series of symmetrically constrained spherical harmonics [5,11] evaluated for a unit vector parallel to the interface, as determined by the field gradient. In contrast, including misorientation-dependence represents a more significant challenge and, consequently, a number of distinct approaches have been attempted.

Perhaps the most widely used approach is the multi-phase-field (MPF) method [2-3,12], which is to simply use a separate scalar phase-field for each grain orientation that appears in the simulation cell [7-8,13]. This scheme unfortunately scales poorly with system size: Both the number of phase fields and the number of differential equations involved grow with the number of grains. While grain remapping algorithms [14-15] mitigate this problem, many authors have sought to devise alternative schemes where a vector-valued field encodes the local grain orientation. Beyond its efficiency, this crystallography-aware representation also facilitates the inclusion of other phenomena, such as elastic effects [16-17] or electrostatic effects [18].

A seminal proposal (the so-called KWC model) [19-21] was to combine a grain orientation field $\theta$ with a scalar order parameter $\eta$ field that tends to one within grains but decreases near GBs. These fields are coupled through a functional that lowers the energy cost of rotating the grain orientation when the order parameter is low, which forces localization of the GB. In this approach, the GB energy is, by necessity, increasing with the magnitude of the misorientation, which prevents the implementation of a fully general misorientation-dependence. This limitation fundamentally arises from the difficulty in determining the orientations of the adjoining grains solely from the field values at a point within a GB.

A modification of this approach [22] has been proposed to remedy this limitation. The idea is to eliminate the constraint that the orientation field $\theta$ be smoothly varying. As a result, in the middle of the GB, the orientation field $\theta$ abruptly jumps from its value in one grain to its value in the other. This ensures that the information regarding both grain orientations is available to calculate the interfacial energy within the GB. A drawback of this scheme is that one must abandon a conventional PDE formulation for the time evolution of the orientation field $\theta$. Instead, the order parameter $\eta$ is evolved smoothly for a small time step, after which the orientation $\theta$ is updated via a thresholding scheme. The whole process involves some non-smooth optimization processes. This approach in principle solves the general orientation-dependence problem, but there remains considerable interest in attempting to achieve the same goal while maintaining conventional smooth time evolution equations.

An alternative approach [6,23-24] is to

maintain smoothly varying orientation fields, but introduce a coupling term between orientation $\theta$ and orientation gradient $\nabla\theta$. The extra explicit dependence on $\theta$ affords additional flexibility that enables the representation of more complex misorientation-dependence. However, there is no mathematical guarantee that any misorientation-dependence can be parameterized in this way. Moreover, the process of determining the free energy functional that implies a given GB energy anisotropy is a complex inverse problem involving a fitting procedure based on numerically equilibrated field profiles.

In this paper, we seek to obtain a compact and efficient set of functional phase-field equations that allow for arbitrary misorientation- and inclination-dependence of interfacial excess free energy, without introducing a large number of auxiliary phase fields. Our approach builds upon these prior insights and further extends the form of free energy functional considered. To simplify the exposition, we initially consider a 2D microstructure where grains can only rotate along one axis (an extension to 3D is described later). The key piece of information that traditional functionals are missing is the orientation of the grains adjoining a given GB. Our proposal specifically addresses this shortcoming by introducing a 'nonlocal' formulation that explicitly incorporates grain orientation changes across the interface.

This paper is organized as follows. Section II introduces the model formulation, key functional components, and the frame invariance analysis. Section III presents the analytical functional derivatives, the asymptotic analysis of the time evolution equation, and the numerical algorithms. In Section IV, we first validate the equilibrium GB profiles, followed by misorientation- and inclination-dependent GB energy. We then demonstrate the model's capability using standard test cases, including GB mobility, Wulff constructions, triple junctions, and polycrystalline systems. Section V gives an overall description of the extensions to 3D systems. Finally, we summarize, discuss the advantages and limitations, and suggest directions for future work.

## II. METHODS

### A. Free Energy Functional

The method takes as input the GB functions $B(\theta^+,\theta^-,\boldsymbol{v})$, which generally depend on both the misorientation, as described by the crystallographic orientation of two adjoining grains (denoted by $\theta^+$ and $\theta^-$), and on the orientation of the GB (i.e., inclination), described by a unit vector $\boldsymbol{v}$. The system domain is divided into three types of regions: 1) inner region, representing points clearly inside a GB, 2) outer region, representing points transition from the GB to the bulk and lying near the bulk lattice, and 3) bulk region, representing points completely inside the bulk lattice. The inner and outer regions together constitute the GB region.

In 2D systems, the free energy functional of GBs is expressed in terms of a single orientation field $\theta(\boldsymbol{x},t) \in \mathbb{R}$, where $\theta(\boldsymbol{x},t)$ is a scalar. It represents the angle between a given crystallographic axis within the grain and the x-axis. The GB energy functional is given by:

$$F = \int (f_1 + f_2)\mathrm{d}\boldsymbol{x}, \tag{2.1}$$

where

$$f_1 = B_{aniso}(\theta^+,\theta^-,\boldsymbol{v})w(|\nabla\theta|)\overline{|\nabla\theta|}^2, \tag{2.2}$$

$$f_2 = B_{iso}(\theta^+,\theta^-)\big(1 - w(|\nabla\theta|)\big)\left[\alpha\overline{|\nabla\theta|}^2 + \beta c\left(\frac{\theta-\theta^-}{\theta^+-\theta^-}\right)\right]. \tag{2.3}$$

In Eq. (2.1), $f_1$ is the inner term that primarily determines the grain shape in the inner region, while $f_2$ is the outer term that controls the GB width. The outer term provides additional smoothness and enforces the bulk orientation constraint by penalizing deviations of $\theta$ from the bulk lattice orientation in the outer region. The parameters $\alpha$ and $\beta$ are constants that control the magnitude of the additional smoothness and the bulk orientation constraint, respectively. When a point lies inside the bulk region, both terms vanish.

In Eqs. (2.2) and (2.3), $|\nabla\theta|$ denotes the magnitude of the orientation field gradient. Quantities $\theta^+$ and $\theta^-$ are the nonlocal orientation fields, which are designed to pick up the bulk lattice orientations on either side of a GB for points locate inside the GB. Accordingly, the misorientation between neighboring grains is given by $\Delta\theta = |\theta^+ - \theta^-|$. The vector $\boldsymbol{v}$ represents the unit normal to the GB.

Terms $f_1$ and $f_2$ involve several auxiliary functions. Here, $B_{iso}(\theta^+,\theta^-)$ is the isotropic GB function, because it depends only on the misorientation and not on the inclination. $B_{aniso}(\theta^+,\theta^-,\boldsymbol{v})$ is the anisotropic GB function, as it depends on both the misorientation and the inclination. Together, these functions determine the misorientation- and inclination-dependence of the GB energy. In Eqs. (2.2) and (2.3), $|\nabla\theta|^2$ is normalized by dividing by the square of the misorientation (i.e., $\overline{|\nabla\theta|}^2 = \frac{|\nabla\theta|^2}{|\theta^+-\theta^-|^2}$). This normalization suppresses the growth of the $|\nabla\theta|^2$ contributions when the misorientation increases, thereby allowing the GB

functions to dominate the misorientation- and inclination-dependent GB energy. The function $w(|\nabla\theta|)$ is a weighting function approaching 1 within the GB and going to 0 away from it, while $c\left(\frac{\theta-\theta^-}{\theta^+-\theta^-}\right)$ is a double-well potential giving preference to orientations $\theta$ close to $\theta^+$ and $\theta^-$. The specific forms of these functions are described in the following subsections. All quantities are designed to vanish in the bulk region.

### B. Nonlocal Orientation Field

Ideally, one would want to 'extrapolate' the value of the orientation field in adjoining grains solely from the knowledge of the orientation field $\theta$ within the GB. FIG 1 illustrates the dependence of $\theta^\pm$ on $\theta$. The values of $\theta^+$ and $\theta^-$ corresponding to a point $x$ within the GB are extrapolated in real space ('$x$-space') as:

$$\theta^\pm(x) = \theta\left(x \pm d\frac{\nabla\theta}{|\nabla\theta|}\right), \tag{2.4}$$

where $d$ is the search distance at which the orientation field transitions from the GB to the bulk. This distance is equal to the equilibrium width of the GB. For 1D systems (i.e. simple slab geometries), $\theta^\pm$ can be reduced to $\theta(x \pm d)$. Note that $\theta^\pm$ can also be extrapolated in '$\theta$-space' (see Appendix A for details). However, for efficiency and simplicity, we consider only the '$x$-space' extrapolation in this paper.

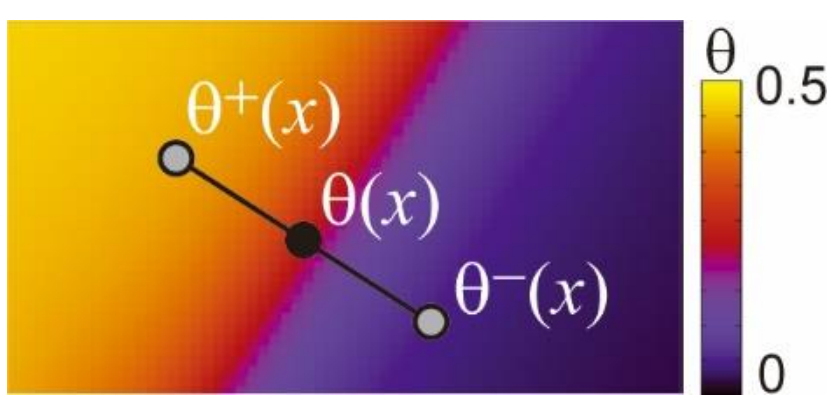

**FIG 1.** A diagram illustrating the 'extrapolation' of the nonlocal orientation fields $\theta^\pm(\boldsymbol{x})$ originating from the local orientation $\theta(\boldsymbol{x})$.

### C. Grain Boundary Functions

GB energy and related properties generally depend on the misorientation between neighboring grains as well as the GB inclination. Accordingly, we aim to construct GB functions as representative forms that capture the essential features of GB energy, including misorientation periodicity and crystallographic inclination anisotropy. These functions are not intended to quantitatively reproduce material-specific GB energies, but rather to provide a flexible and physically consistent framework for modeling GB behavior, allowing systematic investigation of the model behavior independent of specific materials. We choose the following forms:

$$B_{iso}(\theta^+, \theta^-) = |\sin(n\Delta\theta)|, \tag{2.5}$$

$$B_{aniso}(\theta^+, \theta^-, \boldsymbol{v}) = B_{iso}(\theta^+, \theta^-)\left[1 + \epsilon_m \sin\left(m(\theta^* - \Psi)\right)\right]. \tag{2.6}$$

Eq. (2.5) represents the isotropic GB function, while Eq. (2.6) represents the anisotropic GB function. The unit normal vector to the GB is given by $\boldsymbol{v} = \frac{\nabla\theta}{|\nabla\theta|} = [\cos\Psi \quad \sin\Psi]^T$, where $\Psi$ is the GB inclination angle in the reference frame. The parameter $n$ (with $2n \in \mathbb{Z}^+$) represents the symmetry of the crystal lattice, and with the absolute value symbol, $n$ represents $2n$-fold lattice symmetry. The anisotropic coefficient $\epsilon_m \in [0,1)$ controls the strength of the anisotropy, while $m \in \mathbb{Z}^+$ represents the symmetry of the GB inclination. Quantity $\theta^* = \frac{\theta^+ + \theta^-}{2} + K$ represents the GB inclination angle in the material frame, while $K$ is a constant used to adjust preferences for the grain inclination relative to the GB orientation. The reference inclination angle $\Psi$, defined as the angle between the GB normal vector and the x-axis, is given by:

$$\Psi = \begin{cases} \frac{\pi}{2}, & v_1 = 0 \\ \arctan\frac{v_2}{v_1}, & v_1 \neq 0 \end{cases}, \tag{2.7}$$

where $v_1$ and $v_2$ are the $x$ and $y$ components of the GB normal vector, respectively.

As shown in Eq. (2.2), the anisotropic GB function $B_{aniso}(\theta^+, \theta^-, \boldsymbol{v})$ is used to include the anisotropic effects within the inner region. In contrast, anisotropy is not considered in the outer region. Accordingly, as indicated in Eq. (2.3), the isotropic GB function is used in the outer term to ensure the convergence of the GB energy as $|\nabla\theta|$ and the misorientation approach zero. Ideally, the equilibrium GB energy is proportional to a linear combination of the isotropic and anisotropic GB functions.

### D. Weighting Function

As indicated in Eq. (2.1), this model has two contributions. These contributions are intended to act locally on different regions of the system rather than globally. Specifically, when a point $\boldsymbol{x}$ lies within a GB, the inner contribution should dominate, whereas the outer contribution should be suppressed. To achieve this, we introduce a weighting function $w(|\nabla\theta|)$ that indicates whether the current point is clearly within the GB ($w(|\nabla\theta|) = 1$) or clearly in the grains ($w(|\nabla\theta|) = 0$), and transitions smoothly between these values, based on the knowledge of $|\nabla\theta|$:

$$w(|\nabla\theta|) = \begin{cases} 0, & |\nabla\theta| < u_0 \\ \frac{(|\nabla\theta| - u_0)^2(|\nabla\theta| - 2u_1 + u_0)^2}{(u_1 - u_0)^4}, & |\nabla\theta| \in [u_0, u_1], \\ 1, & |\nabla\theta| > u_1 \end{cases} \tag{2.8}$$

where $u_0$ and $u_1$ are constants representing the lower and upper threshold of $|\nabla\theta|$, respectively. Parameter $u_0$ should be close to zero, while $u_1$ can be relatively flexible. Typically, $u_1$ is set to an intermediate value such that $u_0 \ll u_1 < |\nabla\theta|_{max}$. Since $w(|\nabla\theta|)$ is smooth, it must satisfy the following constraints: 1) $w(u_0) = 0$, $w(u_1) = 1$ and 2) $w_u(u_0) = w_u(u_1) = 0$.

Specifically, a point with $u_0 \le |\nabla\theta| \le u_1$ is considered to be entering a GB. If $|\nabla\theta| > u_1$, the point is considered to be fully inside a GB, whereas if $|\nabla\theta| < u_0$, the point is considered to be completely inside the bulk lattice. In the energy functional, the inner contribution is weighted by $w(|\nabla\theta|)$, whereas the outer contribution is weighted by $1 - w(|\nabla\theta|)$. When a point lies inside the GB, the weighting function approaches 1 and the system is dominated by the inner term controlling the grain morphology. Conversely, near the bulk region, the system is governed by the outer term to prevent the orientation field $\theta$ from wandering.

### E. Double-well Potential

The double-well potential $c\left(\frac{\theta-\theta^-}{\theta^+-\theta^-}\right)$ is chosen as:

$$c(u) = u^2(1-u)^2, \tag{2.9}$$

where $u = \frac{\theta-\theta^-}{\theta^+-\theta^-} \in [0,1]$. Quantity $u$ satisfies the following conditions: 1) $u = 0$ when $\theta = \theta^-$ and $u = 1$ when $\theta = \theta^+$; 2) $c(0) = c(1) = c_u(0) = c_u(1) = 0$. For the case $\theta^+ = \theta^-$ (i.e., in the bulk region where $\theta = \theta^+ = \theta^-$), we set $u = 1$ to avoid indeterminate expressions.

This function contributes primarily in the outer region where the orientation field $\theta$ is close to $\theta^+$ or $\theta^-$, as it receives a larger weight there. This is sufficient to ensure that $\theta$ experiences a driving force towards the asymptotic values in the adjoining grains. In the absence of this function, $\theta$ near the bulk region on either side of a GB is unable to remain 'fixed' during time evolution, and the bulk lattice orientations tend to approach each other until the GB completely vanishes.

### F. Frame Invariance

The orientation field $\theta$ is not a frame-invariant quantity by itself, since its value changes when observed from different reference frames. Nevertheless, this model is inherently frame-invariant, despite the inclusion of bulk orientation terms.

Assume that the system is observed from a reference frame rotated by an angle $\delta\theta$ relative to the original frame. Under this rigid rotation, the orientation-related quantities transform as:

$$\theta \to \theta + \delta\theta,\ \ \theta^+ \to \theta^+ + \delta\theta,\ \ \theta^- \to \theta^- + \delta\theta. \tag{2.10}$$

Similarly, the other angular quantities transform as:

$$\theta^* \to \theta^* + \delta\theta, \qquad \Psi \to \Psi + \delta\theta. \tag{2.11}$$

In contrast, all gradient-related quantities remain invariant under such a rigid rotation:

$$\nabla\theta \to \nabla\theta, \qquad |\nabla\theta| \to |\nabla\theta|. \tag{2.12}$$

Substituting these transformations into Eq. (2.1) shows that the energy functional remains unchanged, thereby demonstrating the frame invariance of this model.

## III. SOLUTIONS

### A. Time Evolution Equation

The GB system is a typical non-conservative field. Therefore, we apply the Allen-Cahn type time evolution equation [25-27]:

$$\frac{\partial\theta}{\partial t} = -M\frac{\delta F}{\delta\theta}, \tag{3.1}$$

where $M$ is the mobility of the orientation field and may depend on misorientation and inclination. In this model, the orientation mobility $M$ is taken to be constant for simplicity. The rationale for this choice is discussed in Section III B and Appendix B. $\frac{\delta F}{\delta\theta}$ can be analytically calculated by the calculus of variations:

$$\frac{\delta F}{\delta\theta} = \frac{\partial f}{\partial\theta} - \nabla\cdot\frac{\partial f}{\partial(\nabla\theta)}, \tag{3.2}$$

where $f = f_1 + f_2$.

Beyond Eq. (3.1), the full time evolution equations also include the update of the 'lookup direction' based on $\nabla\theta$, in the definition of $\theta^\pm(\boldsymbol{x})$ (Eq. (2.4)). However, the functional derivative (Eq. (3.2)) purposely does not include the effect of the rotation of the gradient $\nabla\theta$ in Eq. (2.4) because this effect can be shown to be zero to first order. To see this, consider three cases. Near a grain boundary, the sets of constant value of $\theta$ are parallel to the grain boundary while $\nabla\theta$ is perpendicular to it. Therefore, for infinitesimal rotations of $\nabla\theta$, points of the form $\boldsymbol{x} \pm d\frac{\nabla\theta}{|\nabla\theta|}$ will remain on a set of constant $\theta$ and the changes in $\theta^\pm$ are zero to first order. Within a grain or near grain

boundary junction, our functional's dependence on $\theta^{\pm}$ is turned off through the weight function $w(|\nabla\theta|)$ going to zero.

Using the simple notations $B_{iso}$, $B_{aniso}$, $w$, and $c$ for the functions Eq. (2.5), Eq. (2.6), Eq. (2.8), and Eq. (2.9), respectively, the derivatives in Eq. (3.2) are given by:

$$\frac{\partial f}{\partial \theta} = B_{iso}\beta(1-w)\frac{\partial c}{\partial \theta}, \tag{3.3}$$

$$\begin{aligned}\frac{\partial f}{\partial(\nabla\theta)} &= \frac{1}{|\theta^+-\theta^-|^2}\left[B_{aniso}\cdot\left(2w\nabla\theta+\frac{\partial w}{\partial(\nabla\theta)}|\nabla\theta|^2\right) + \frac{\partial B_{aniso}}{\partial(\nabla\theta)}w|\nabla\theta|^2\right] \\ &+ \frac{B_{iso}\alpha}{|\theta^+-\theta^-|^2}\left[2(1-w)\nabla\theta - \frac{\partial w}{\partial(\nabla\theta)}|\nabla\theta|^2\right] \\ &- B_{iso}\beta\frac{\partial w}{\partial(\nabla\theta)}c. \end{aligned} \tag{3.4}$$

The derivatives of functions $c$, $B_{aniso}$, and $w$ appearing in Eq. (3.3) and Eq. (3.4) are:

$$\frac{\partial c}{\partial \theta} = 2\frac{\theta-\theta^-}{(\theta^+-\theta^-)^2}\left[1-3\left(\frac{\theta-\theta^-}{\theta^+-\theta^-}\right) + 2\left(\frac{\theta-\theta^-}{\theta^+-\theta^-}\right)^2\right], \tag{3.5}$$

$$\begin{aligned}\frac{\partial B_{aniso}}{\partial(\nabla\theta)} &= \mathbf{0}, \qquad if\ |\nabla\theta| < u_0, \\ \frac{\partial B_{aniso}}{\partial(\nabla\theta)} &= |\sin[n(\theta^+-\theta^-)]| \times \frac{m\epsilon_m\cos[m(\theta^*-\Psi)]}{|\nabla\theta|^2}\boldsymbol{U}\nabla\theta, \ \ if\ |\nabla\theta| \geq u_0,\end{aligned} \tag{3.6}$$

$$\begin{aligned}\frac{\partial w}{\partial(\nabla\theta)} &= \mathbf{0}, \qquad if\ |\nabla\theta| < u_0 \cup |\nabla\theta| > u_1, \\ \frac{\partial w}{\partial(\nabla\theta)} &= (|\nabla\theta|-u_0)(|\nabla\theta|-2u_1+u_0)(|\nabla\theta|-u_1) \times \frac{4}{(u_1-u_0)^4}\frac{\nabla\theta}{|\nabla\theta|}, \qquad if\ |\nabla\theta| \in [u_0,u_1],\end{aligned} \tag{3.7}$$

where $\boldsymbol{U} = \begin{bmatrix} 0 & 1 \\ -1 & 0 \end{bmatrix}$ is a unitary that converts a vector to its normal. Note that for 2D systems, $\frac{\partial c}{\partial \theta} \in \mathbb{R}$ and $\frac{\partial B}{\partial(\nabla\theta)}, \frac{\partial w}{\partial(\nabla\theta)} \in \mathbb{R}^2$.

### B. Asymptotic Analysis

Although it is unlikely that an analytical solution for such a highly nonlinear PDE can be obtained, important properties of the equilibrium solutions can still be obtained through an asymptotic analysis.

Referring to Appendix B for details, the asymptotic analysis recovers the classical grain-growth relation $V_n = -M_{GB}\gamma\kappa$, where $V_n$ is the normal GB velocity, $\kappa$ is the GB curvature, and $M_{\mathrm{GB}}$ is the GB mobility. The GB mobility is related to the orientation field mobility $M$ via:

$$M_{GB} = M\frac{\int_{-\infty}^{+\infty} h_0\frac{\partial\theta_0}{\partial\eta}\mathrm{d}\eta}{\gamma\Delta\theta}. \tag{3.8}$$

Here, $\eta$ is the scaled coordinate normal to the GB, and $\gamma$ is the GB energy density, which can be computed numerically. The field $\theta_0$ denotes the leading-order solution of Eq. (3.1), and $h_0$ denotes the zeroth-order term of a scalar function $h(\nabla\theta,|\nabla\theta|)$ obtained from Eq. (3.4). Consequently, the GB mobility can be computed by substituting the prescribed orientation field mobility and the numerically obtained GB profiles into the quantities on the right side of Eq. (3.8).

### C. Numerical Solutions

The PDE given in Eq. (3.1) can be solved using various numerical methods, including the finite difference method (FDM), finite volume method (FVM), and finite element method (FEM). In this section, we employ the FDM, while a discussion of the FEM implementation is deferred to the Discussion section. One principal motivation for using the FDM is that the free energy functional involves explicit dependence on nonlocal orientation fields, for which the spatially uniform discretization of FDM enables straightforward extrapolation of nonlocal sampling points without introducing additional auxiliary functional kernels. Moreover, the FDM formulation provides sufficient numerical precision and naturally adapts to commonly used boundary conditions in grain growth simulations, such as periodic and zero-flux boundary conditions.

We use second-order central differences for spatial discretization and the explicit forward Euler scheme for time integration. Let $\theta_{i,j}^k$ denote the value of $\theta$ at grid node $(i,j)$ and time step $k$. The discrete gradient and time derivative are approximated as:

$$\nabla\theta_{i,j}^k = \begin{bmatrix} \dfrac{\theta_{i+1,j}^k-\theta_{i-1,j}^k}{2\Delta x} \\ \dfrac{\theta_{i,j+1}^k-\theta_{i,j-1}^k}{2\Delta y} \end{bmatrix}, \qquad \frac{\partial\theta_{i,j}^k}{\partial t} = \frac{\theta_{i,j}^{k+1}-\theta_{i,j}^k}{\Delta t}, \tag{3.9}$$

where $\Delta x$ and $\Delta y$ are the grid spacings in the $x$ and $y$ directions, respectively, and $\Delta t$ is the time step size.

The divergence term $\nabla \cdot \frac{\partial f}{\partial(\nabla\theta)}$ in Eq. (3.2) is computed numerically. Given the complexity of Eq. (3.4), computing $\nabla \cdot \frac{\partial f}{\partial(\nabla\theta)}$ analytically is extremely troublesome. Moreover, the divergence operator $(\nabla \cdot)$ introduces higher order of powers of $|\nabla\theta|$ in the denominator, which can cause numerical instability when the misorientation (and thus $|\nabla\theta|$) is small. We therefore compute this term using FDM as well.

Substituting Eq. (3.9) into Eq. (3.4), we obtain the discrete gradient derivative $\left[\frac{\partial f}{\partial(\nabla\theta)}\right]_{i,j}^{k}$, which is written as $\left[\frac{\partial f}{\partial(\nabla\theta)}\right]_{i,j}^{k} = \begin{bmatrix} f_{\nabla\theta_{i,j},x}^{k} \\ f_{\nabla\theta_{i,j},y}^{k} \end{bmatrix}$. The corresponding discrete divergence term $\left[\nabla \cdot \frac{\partial f}{\partial(\nabla\theta)}\right]_{i,j}^{k}$ is then given by:

$$\left[\nabla \cdot \frac{\partial f}{\partial(\nabla\theta)}\right]_{i,j}^{k} = \frac{f_{\nabla\theta_{i+1,j},x}^{k} - f_{\nabla\theta_{i-1,j},x}^{k}}{2\Delta x} + \frac{f_{\nabla\theta_{i,j+1},y}^{k} - f_{\nabla\theta_{i,j-1},y}^{k}}{2\Delta y}. \quad (3.10)$$

Substituting Eq. (3.10) into Eq. (3.1), we obtain the discrete Allen-Cahn equation of motion:

$$\theta_{i,j}^{k+1} = \theta_{i,j}^{k} - \Delta t M \times \left[ f_{\theta_{i,j}}^{k} - \left( \frac{f_{\nabla\theta_{i+1,j},x}^{k} - f_{\nabla\theta_{i-1,j},x}^{k}}{2\Delta x} + \frac{f_{\nabla\theta_{i,j+1},y}^{k} - f_{\nabla\theta_{i,j-1},y}^{k}}{2\Delta y} \right) \right], \quad (3.11)$$

where $f_{\theta_{i,j}}^{k}$ denotes the discrete bulk derivative $\left(\frac{\partial f}{\partial\theta}\right)_{i,j}^{k}$ obtained by substituting Eq. (3.9) into Eq. (3.3).

### D. Algorithm

In this section, we present the details of the implementation of the extrapolation for $\theta^{\pm}$. A gradient search algorithm is developed to extrapolate $\theta^{\pm}$ in the $\boldsymbol{x}$-space, as defined in Eq. (2.4). This algorithm is robust and applicable for both GBs and triple junctions.

We begin by describing its application to GBs. **FIG 2** illustrates the gradient search algorithm for a given local point $X$. If the point lies inside a GB (i.e., $|\nabla\theta| \geq u_0$), we first determine the unit GB normal vector $\boldsymbol{v}$. The magnitudes of the components of this normal vector are used as the search step sizes in the $x$ and $y$ directions. At each step, we evaluate whether $|\nabla\theta|$ remains greater than $u_0$. Once $|\nabla\theta| < u_0$, the search is terminated, indicating that the bulk region has been reached, and the value of $\theta$ at the stopping point $X'$ is recorded. The same procedure is then repeated along the opposite normal direction $-\boldsymbol{v}$, yielding a second stopping point $X''$. The extrapolated values $\theta^{\pm}$ at point $X$ are thus obtained as $\theta(X')$ and $\theta(X'')$. Without loss of generality, the larger value is assigned to $\theta^{+}$ and the smaller one to $\theta^{-}$. If the point $X$ does not lie inside a GB (i.e., $|\nabla\theta| < u_0$), the procedure becomes simpler. In this case, both $\theta^{\pm}(X)$ are set equal to $\theta(X)$, ensuring that the GB energy vanishes in the bulk region.

For multi-grain systems, the treatment of triple junctions requires special attention [6,22,28-29]. Unlike the GBs, where the extrapolation directions for $\theta^{\pm}$ are well defined, the locations of $\theta^{\pm}$ at triple junctions are inherently ambiguous. Nevertheless, the gradient search algorithm can still be applied with appropriate modifications.

At GBs, searches along the two opposite normal directions always terminate in the bulk within a bounded distance. In contrast, at a triple junction, three GBs intersect but only two opposite search directions are available. As a result, if one search direction is perpendicular to one GB, its opposite direction may be nearly parallel to another GB, causing the search to remain within the GB over a long distance. To handle this situation, we introduce a maximum search distance, beyond which the search is forced to terminate. This distance can be chosen to be on the order of the GB width. If the search starting from point $X$ along one direction reaches this maximum distance while $|\nabla\theta| \geq u_0$ remains satisfied, point $X$ is identified as belonging to a triple junction. In this case, $\theta^{\pm}(X)$ are assigned as the values of $\theta$ at the terminating points $X'$ and $X''$, even if these points are not in the bulk region. This treatment ensures the continuity of misorientation fields within non-bulk regions and prevents numerical abnormalities in long-time simulations.

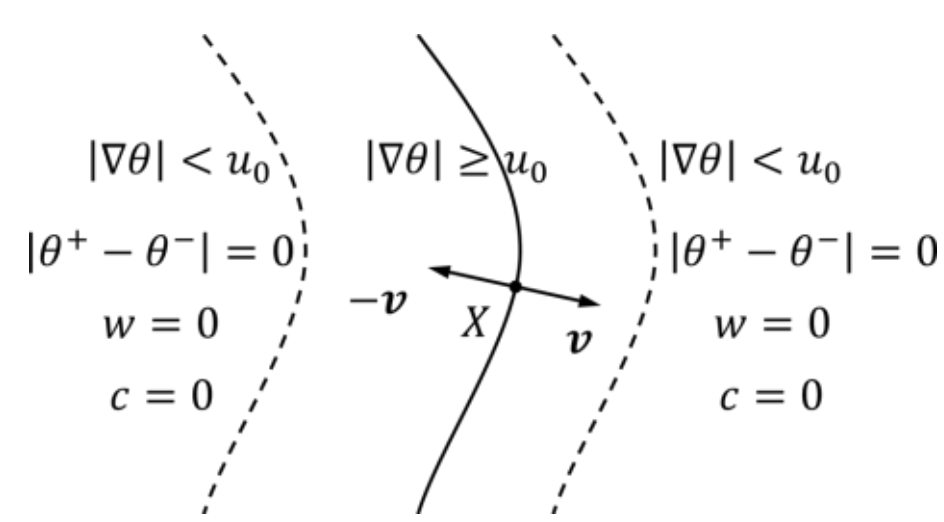

**FIG 2.** A schematic diagram of the gradient search algorithm. The solid curve represents a contour line inside a GB, while the dashed lines represent the two edges of the GB. The arrows indicate two opposite GB normal vectors at point $X$ on the contour line.

### E. Simulation Parameters

We divide all the parameters into two groups: 1) Integration parameters (i.e., $N_x, N_y, \Delta x, \Delta y, \Delta t, err$); 2)

Material parameters (i.e., $M, \alpha, \beta, u_0, u_1, n, m, \epsilon_m, K$). Material parameters $M$, $\alpha$, $\beta$, $n$, $m$, and $\epsilon_m$ will be tuned in the Results section. All parameters are considered dimensionless.

Except otherwise specified, $N_x = 128$ and $N_y = 128$, representing the number of grid points in $x$ and $y$ directions, respectively. The grid spacing is $\Delta x = \Delta y = 0.1$. The time step length is $\Delta t = 10^{-5}$. The converging criterion for the energy is denoted by the error threshold $err$, and $err = 10^{-6}$ provides both sufficient accuracy and computing efficiency. The following material parameters are used throughout the paper unless explicitly specified: $M = 1$, $\alpha = 0.01$, $\beta = 1$, $u_0 = 0.0001$, $u_1 = 32$, $n = 2$, $m = 4$, $\epsilon_4 = 0$, and $K = 0$. Note that we have written $\epsilon_m$ as $\epsilon_4$ for $m = 4$. Periodic boundary conditions are applied to all simulations.

The parameters used in this paper were selected to yield convergent time integrations. However, if the readers wish to select their own parameters, the convergence condition based on the Courant-Friedrichs-Lewy (CFL) criterion [30] must be satisfied. Below, we provide the Courant numbers for terms $f_1$ and $f_2$:

$$C_{f_1} \sim \Delta t M \left( \frac{1}{\Delta x^2} + \frac{1}{\Delta y^2} \right),$$

$$C_{f_2} \sim \max\left\{ \Delta t M \alpha \left( \frac{1}{\Delta x^2} + \frac{1}{\Delta y^2} \right), M\beta\Delta t \right\}. \qquad (3.12)$$

Substituting the parameter values into Eq. (3.12), we obtain $C_{f_1} \sim 2 \times 10^{-3}$ and $C_{f_2} \sim 2 \times 10^{-5}$. Both of them are far less than $C_{max}$, which is typically 1.

## IV. RESULTS

### A. Equilibrium Grain Boundary Profiles

Quasi-1D systems (**FIG 3**) are used to simulate GB profiles. For computational efficiency, the simulation domain is set to $N_x = N_y = 64$. **FIG 4(a)** shows the time evolution of the GB profile. Owing to symmetry in the $x$ dimension, only left half of the GB profile is plotted. The gray curve represents the initial sharp interface. As the simulation continues, the GB becomes smooth and converges toward the analytical steady-state solution (black dashed curve, see Appendix A for the derivation). The simulation results show excellent agreement with the analytical solution. Moreover, the evolution reaches equilibrium rapidly, within approximately 0.5 time units.

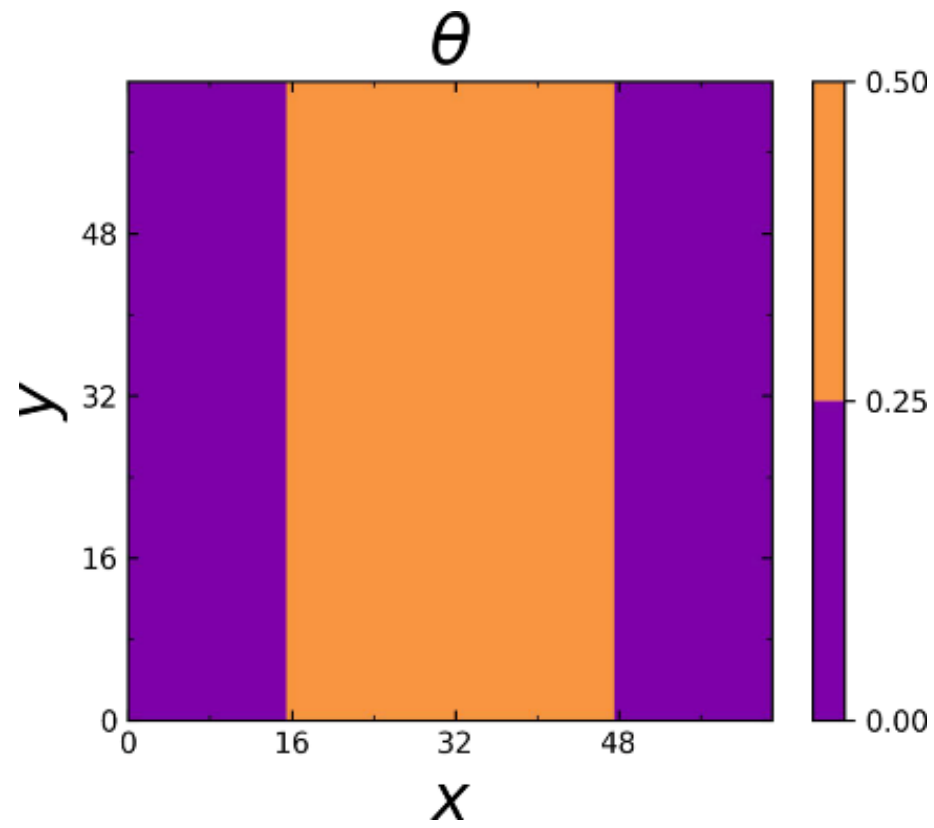

**FIG 3.** The contour plot of a quasi-1D simulation system.

**FIG 4(b)** shows the simulation results for various equilibrium GB profiles with different misorientations $\Delta\theta$. The blue and black dashed curves represent the extrapolated $\theta^+$ and $\theta^-$ solutions, respectively, for $\Delta\theta = 0.75$. The window enclosed by the $\theta^\pm$ curves outlines the GB region, including both the inner and outer regions. In addition, the equilibrium GB width is observed to be nearly independent of misorientation.

The equilibrium GB width is nevertheless tunable and is primarily controlled by the material parameters $\alpha$ and $\beta$ in Eq. (2.3). As shown in **FIGs 4(c)** and **4(d)**, increasing $\alpha$ smoothens the GB profile, resulting in wider interfaces, whereas increasing $\beta$ sharpens the GB profile by strengthening the bulk orientation constraint.

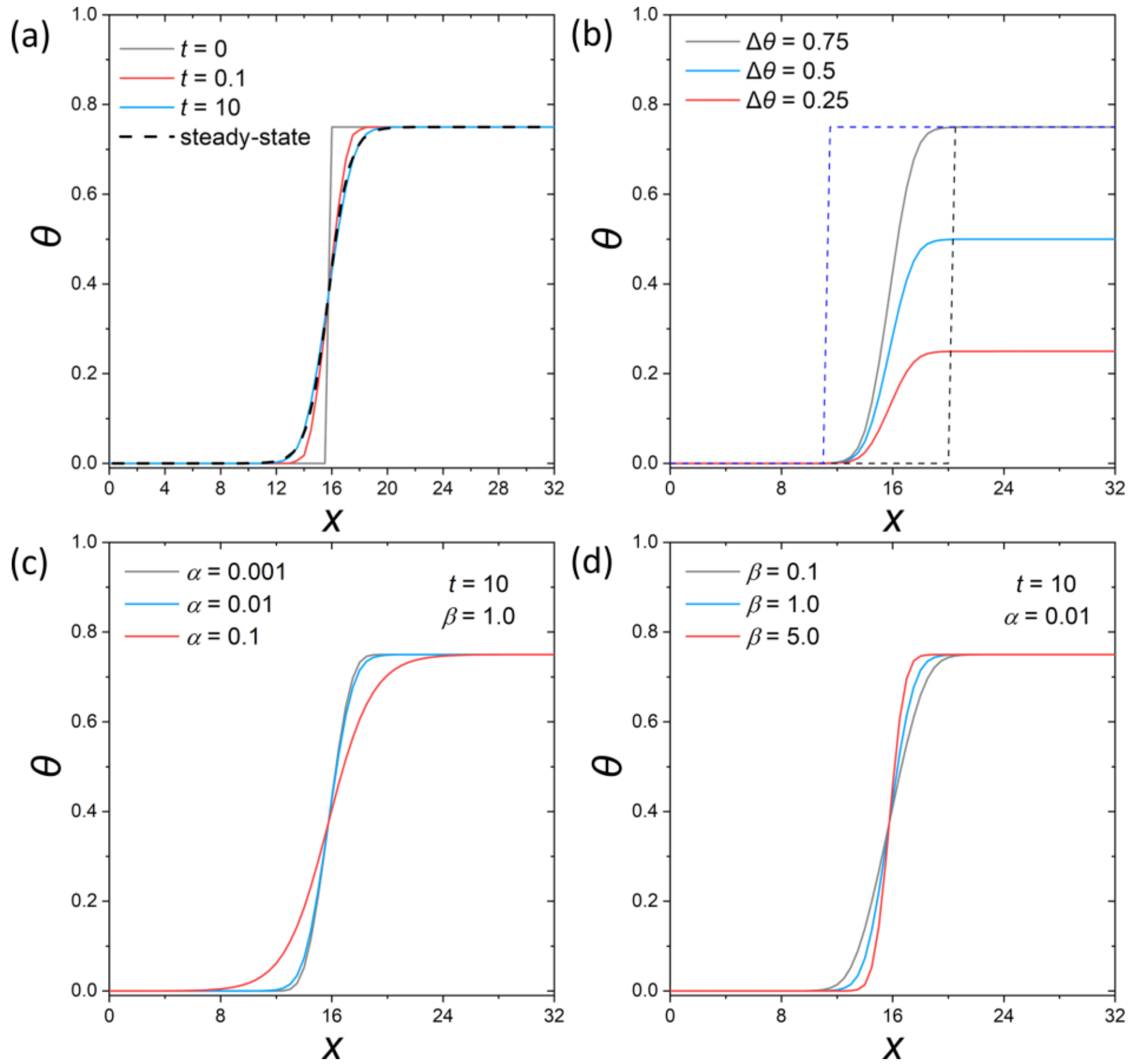

**FIG 4.** (a) Time evolution of the GB profiles for $\Delta\theta = 0.75$. (b) Equilibrium GB profiles for different misorientations obtained from simulations (solid curves), together with the extrapolated $\theta^{\pm}$fields determined using the gradient-search algorithm (dashed curves). Effects of parameters (c) $\alpha$ and (d) $\beta$ on the smoothing and sharpening behavior of the equilibrium GB profiles.

### B. Grain Boundary Energy

The GB energy dependence calculated by the classical KWC model is restricted to the Read-Shockley type [31-32], which is a monotonically increasing function of misorientation. However, this contrasts with both experimental [33] and computational [34-35] observations. Motivated by these discrepancies, we validate in this section how the proposed GB functions $B_{aniso}$ and $B_{iso}$ control the equilibrium GB energy dependence.

The total GB energy $F$ is computed numerically using Eq. (2.1) with simulation systems similar to that shown in **FIG 3**. The time evolution of the total GB energy is shown in **FIG 5**, with results recorded every 1000 time steps. When the energy difference between consecutive outputs falls below the error threshold $err$, the system is considered to have reached equilibrium, and the final recorded value is taken as the equilibrium GB energy. To eliminate finite-size effects, we use the GB energy density in subsequent calculations, given by:

$$\gamma = \frac{F}{2L_y}, \tag{4.1}$$

where $L_y = N_y\Delta y$ is the GB length. All computed values of $\gamma$ are then compared with a linear combination of the isotropic and anisotropic GB functions:

$$aB_{iso} + bB_{aniso} = (a+b)|\sin(n\Delta\theta)| \times \left[1 + \frac{b}{a+b}\epsilon_4 \sin\big(m(\theta^* - \Psi)\big)\right], \tag{4.2}$$

with $a = 0.036$ and $b = 0.006$. These coefficients are 'global' fitting constants and remain unchanged throughout this paper.

**FIG 6** presents the misorientation-dependent GB energy density for isotropic systems ($\epsilon_4 = 0$), in which the misorientation is adjusted by fixing $\theta^- = 0$ and varying $\theta^+$. Symbols denote the simulation results, while solid curves correspond to the combined GB functions given by Eq. (4.2). The effect of lattice symmetry is also illustrated in **FIG 6**. The blue symbols denote two-fold symmetry ( $n = 1$ ), exhibiting a periodicity of $\pi$, while the red symbols correspond to four-fold symmetry ($n = 2$), with a periodicity of $\frac{\pi}{2}$. Using the four-fold symmetry case, a GB with a smaller misorientation ($\Delta\theta < \frac{\pi}{4}$) has the same energy as one with the complementary larger misorientation ($\frac{\pi}{2} - \Delta\theta$). Thus, GBs with complementary misorientations are physically equivalent. **FIG 7** illustrates the equilibrium profiles of two equivalent GBs with complementary misorientations. However, previous studies have reported that nonphysical topological defects [6,24,36] may arise when these GBs meet, since there is no continuous transformation between them. To prevent such defects, the initial misorientation is restricted to the smaller values, specifically $\Delta\theta < \frac{\pi}{2n}$.

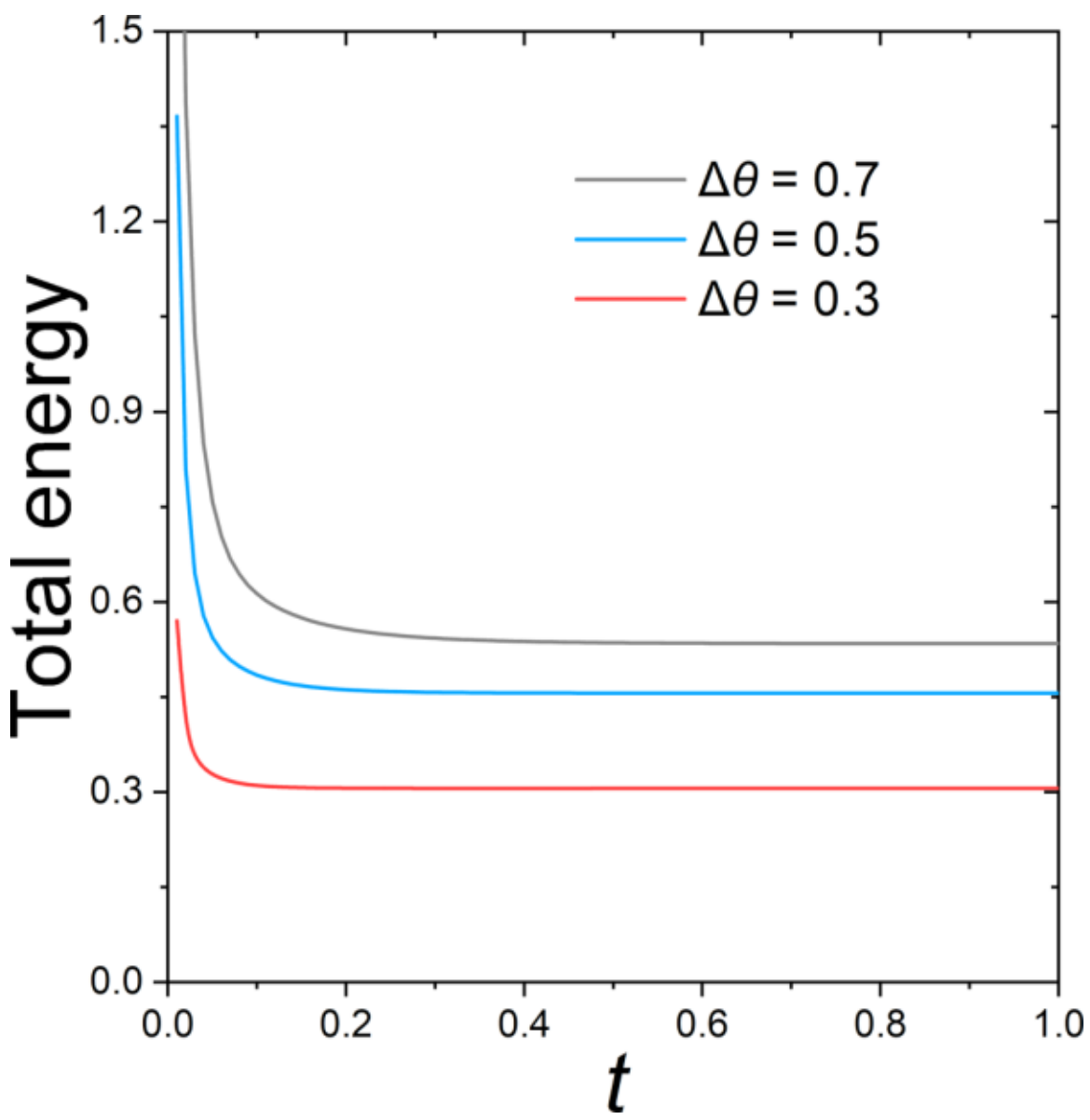

**FIG 5.** Time evolution of the GB energy for different misorientations.

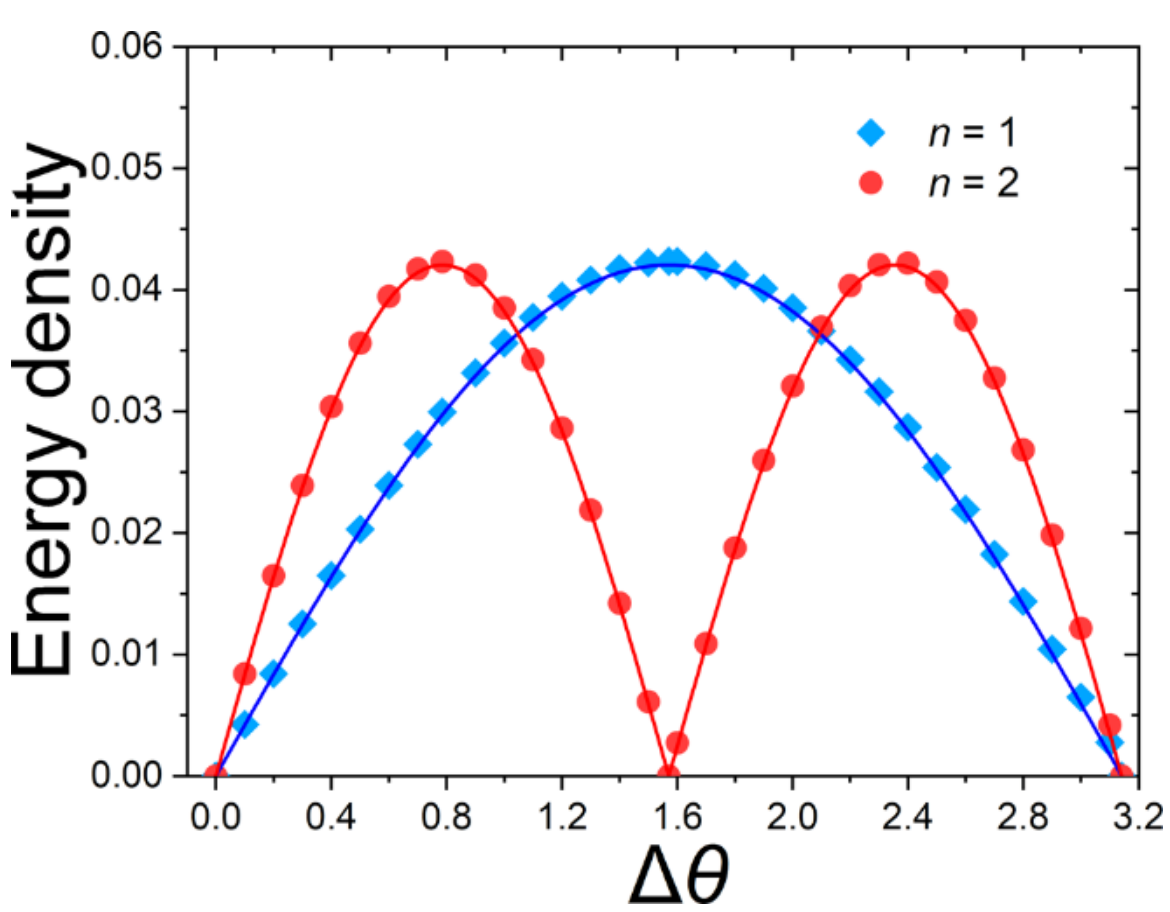

**FIG 6.** Misorientation-dependent GB energy density for isotropic models. Symbols represent simulation results, while solid curves correspond to the the combined GB function $(a + b)|\sin(n\Delta\theta)|$, where $n = 1$ for the blue curve and $n = 2$ for the red curve.

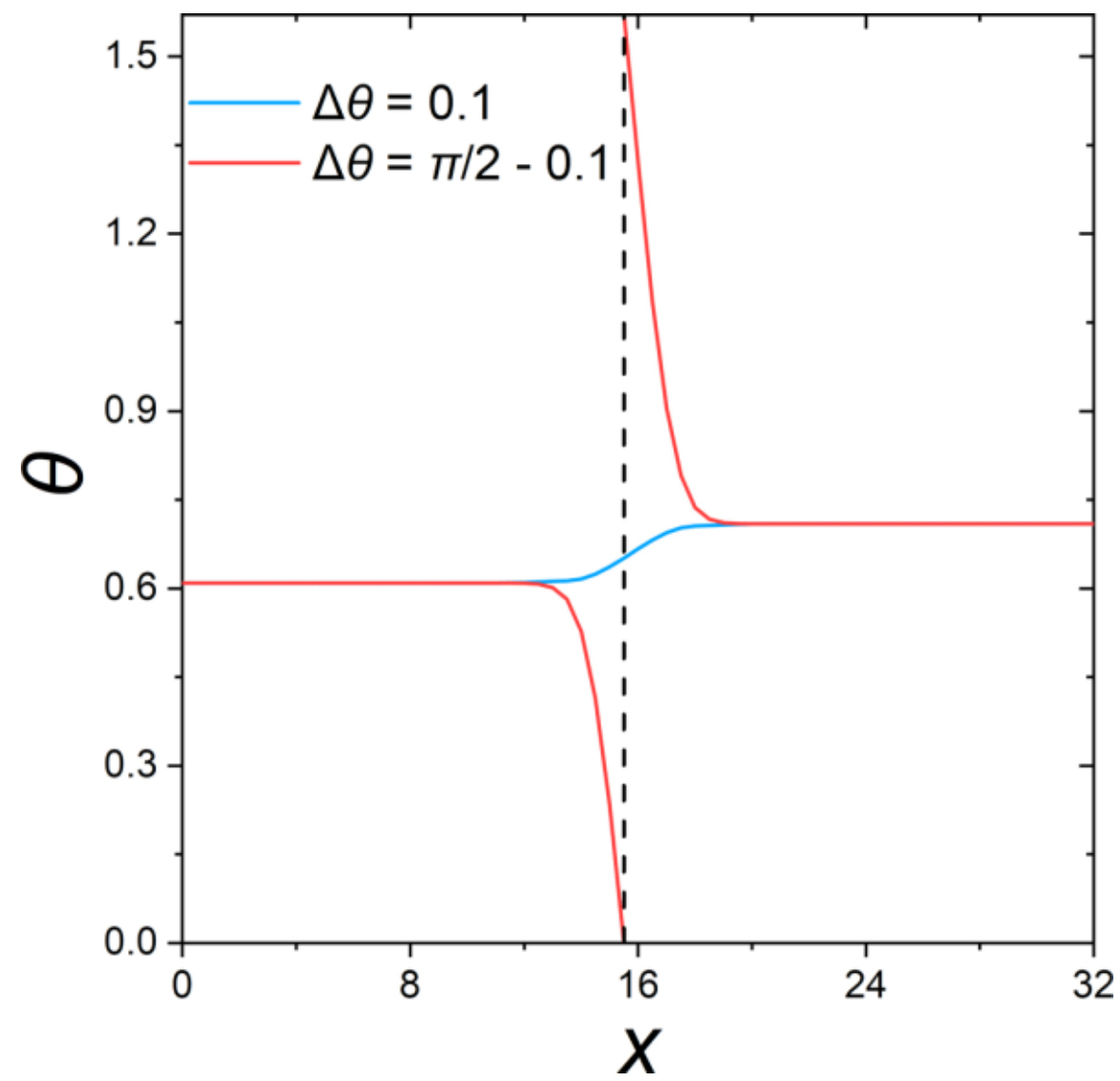

**FIG 7.** Equilibrium GB profiles for two complementary misorientations. $\theta = 0$ is equivalent to $\theta = \frac{\pi}{2}$.

We next incorporate anisotropy and perform the same set of simulations as in **FIG 6** for a system with four-fold inclination symmetry. As indicated by the blue symbols in **FIG 8**, in contrast to the isotropic case, a difference in the amplitudes of the two arches emerges due to the presence of anisotropy. To reverse the relative amplitudes of the two arches, we fix $\theta^+ = 0$ and vary $\theta^-$ when adjusting the misorientation. The corresponding results are indicated by the red symbols in **FIG 8**. The resulting difference in amplitudes reflects the approach used to vary the misorientation. In this approach, changes in $\Delta\theta$ simultaneously modify the quantity $\theta^* = \frac{\theta^+ + \theta^-}{2}$, which corresponds to different inclinations in the material frame.

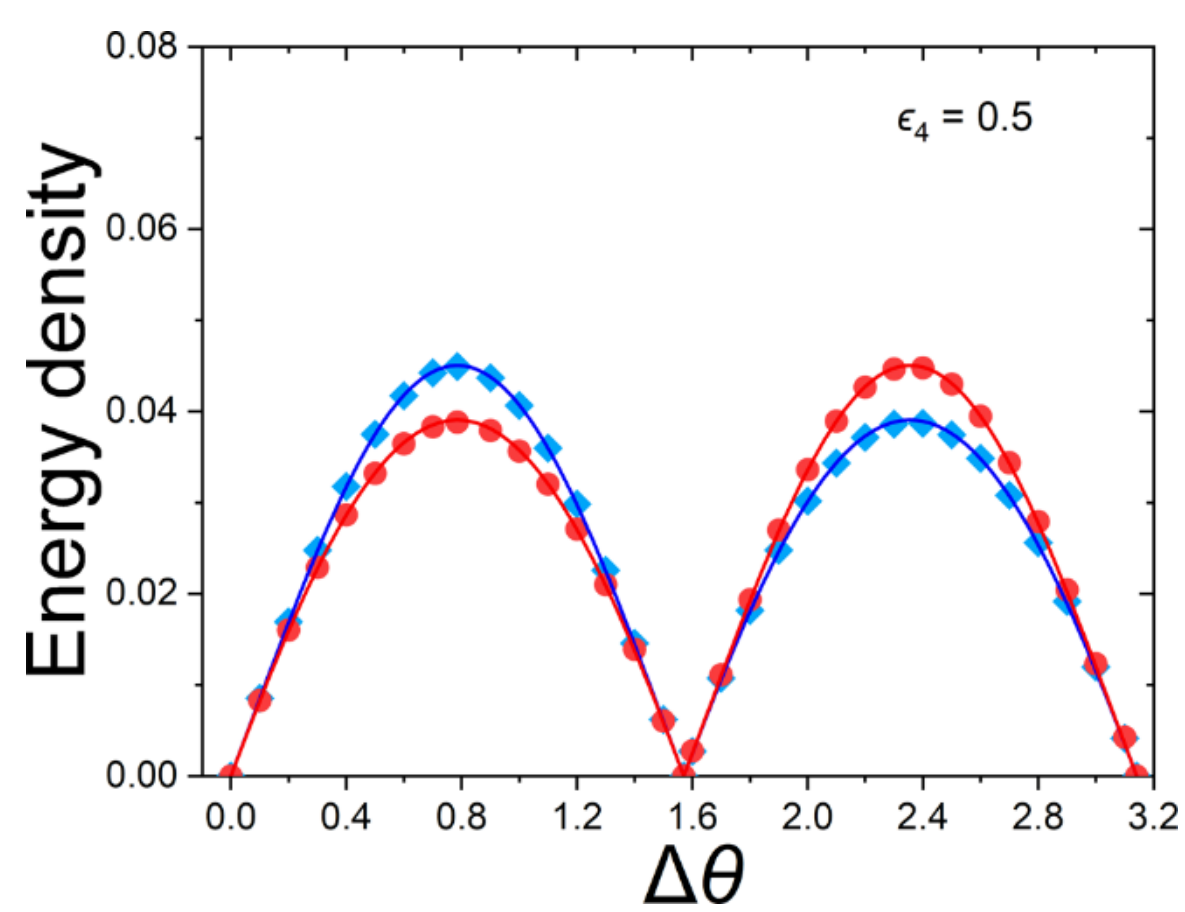

**FIG 8.** Misorientation-dependent GB energy density

for anisotropic systems. Symbols represent simulation results, and solid curves correspond to the combined GB function given by $(a+b)|\sin(2\Delta\theta)|\left[1+\frac{b}{a+b}\epsilon_4\sin(4\theta^*)\right]$. Blue symbols correspond to fixing $\theta^-=0$ and varying $\theta^+$, whereas red symbols correspond to fixing $\theta^+=0$ and varying $\theta^-$.

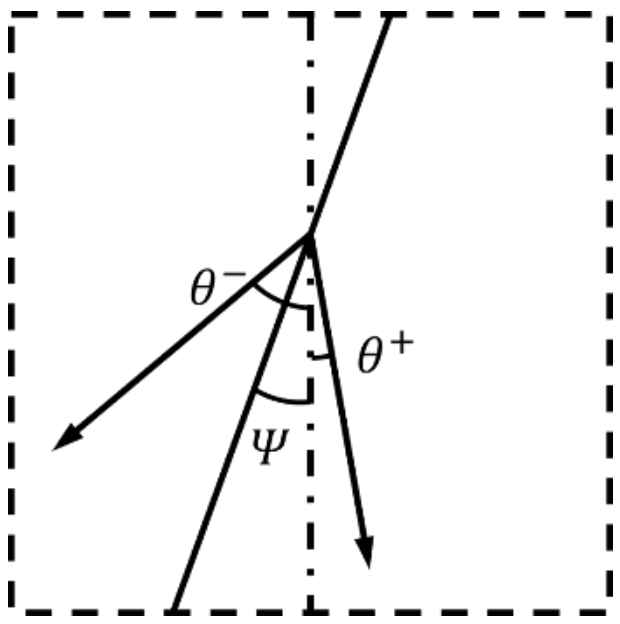

**FIG 9.** A schematic diagram of the reference frame (dashed-dotted line) and the material frame (solid line), with arrows indicating the orientations of the two grains.

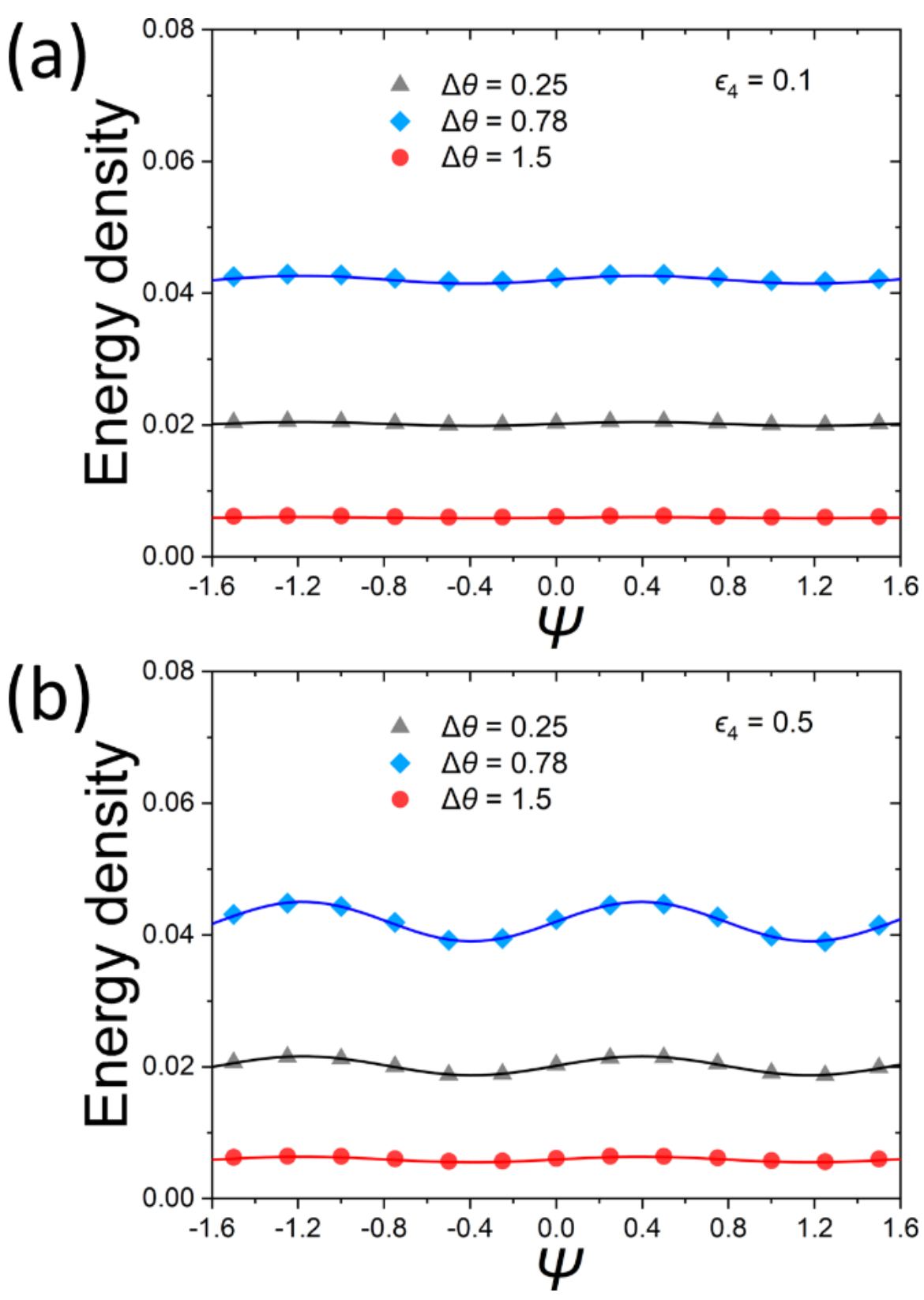

**FIG 10.** Inclination-dependent GB energy for different misorientations with (a) $\epsilon_4=0.1$ and (b) $\epsilon_4=0.5$. Symbols represent simulation results, while solid curves correspond to the combined GB function given by $(a+b)|\sin(2\Delta\theta)|\left[1+\frac{b}{a+b}\epsilon_4\sin(4\Psi)\right]$.

When varying the inclinations, it is important to preserve the intended symmetry of the system. Maintaining a fixed GB geometry avoids GB bending or any other irregularity near the edges of the simulation domain, thereby ensuring accurate computation of the GB energy density. **FIG 9** illustrates the method used to adjust the inclination without altering the GB geometry. The GB (dash-dotted line) is taken as the reference frame and remains fixed throughout the simulation, while the inclination is varied only in the material frame by rigidly rotating the two grains at a fixed misorientation. The inclination angle $\Psi$ is then equivalent to the angle between the GB and the bisector (solid line) of the two grain orientations. **FIG 10** shows the inclination-dependent GB energy density for three different misorientations. The anisotropic coefficient is chosen as 0.1 for **FIG 10(a)** and 0.5 for **FIG 10(b)**. The GB energy density dependence exhibits a sinusoidal form, with the amplitude positively correlated to the energy density level and the anisotropy.

All simulation results in **FIGs 6**, **8**, and **10** are in excellent agreement with Eq. (4.2), demonstrating that the proposed GB functions enable precise control of the misorientation- and inclination-dependent GB energy.

### C. Grain Boundary Mobility

The simulations of GB mobility $M_{GB}$ are performed using an isotropic grain growth system in which a shrinking circular grain is embedded within a large surrounding grain. In this configuration, GB motion is driven only by the local curvature of the GB. It has been reported that the change in the radius of the shrinking grain is given by [37-38]:

$$R_0^2-R^2=Kt, \tag{4.3}$$

where $R_0$ and $R$ are initial and current grain radii of the circular grain, respectively, $t$ is the time, and $K$ is a temperature-dependent constant given by the Arrhenius' equation [39]. Note that Eq. (4.3) can also be derived from the classical relationship $V_n=-M_{GB}\gamma\kappa$, which yields $M_{GB}=\frac{K}{2\gamma}$.

The advantage of this simulation system is notable. Because the radial GB profile $\theta(r)$, the GB energy density $\gamma$, and the misorientation $\Delta\theta$ in Eq. (3.8) remain constant throughout the simulation, the GB mobility $M_{GB}$ has a linear dependence on the orientation mobility $M$. Consequently, although $M_{GB}$ cannot be set directly, it can be easily controlled through the choice of $M$, yielding $M_{GB}\propto M$.

**FIG 11** illustrates the time evolution of the circular grain fraction for two orientation mobilities,

$M = 1$ and $M = 2$. The initial radius is set to $40\Delta x$, and the shrinkage of the circular grain is depicted as the inset of **FIG 11**. The simulation results display an excellent linear dependence on time, in good agreement with Eq. (4.3). This linear relationship further validates that the GB mobility $M_{GB}$ remains constant during the simulation as well as the dependence $M_{GB} \propto M$. The fitted slopes are 0.00581 for $M = 1$ and 0.0109 for $M = 2$, indicating that the shrinking rate is approximately proportional to the GB mobility.

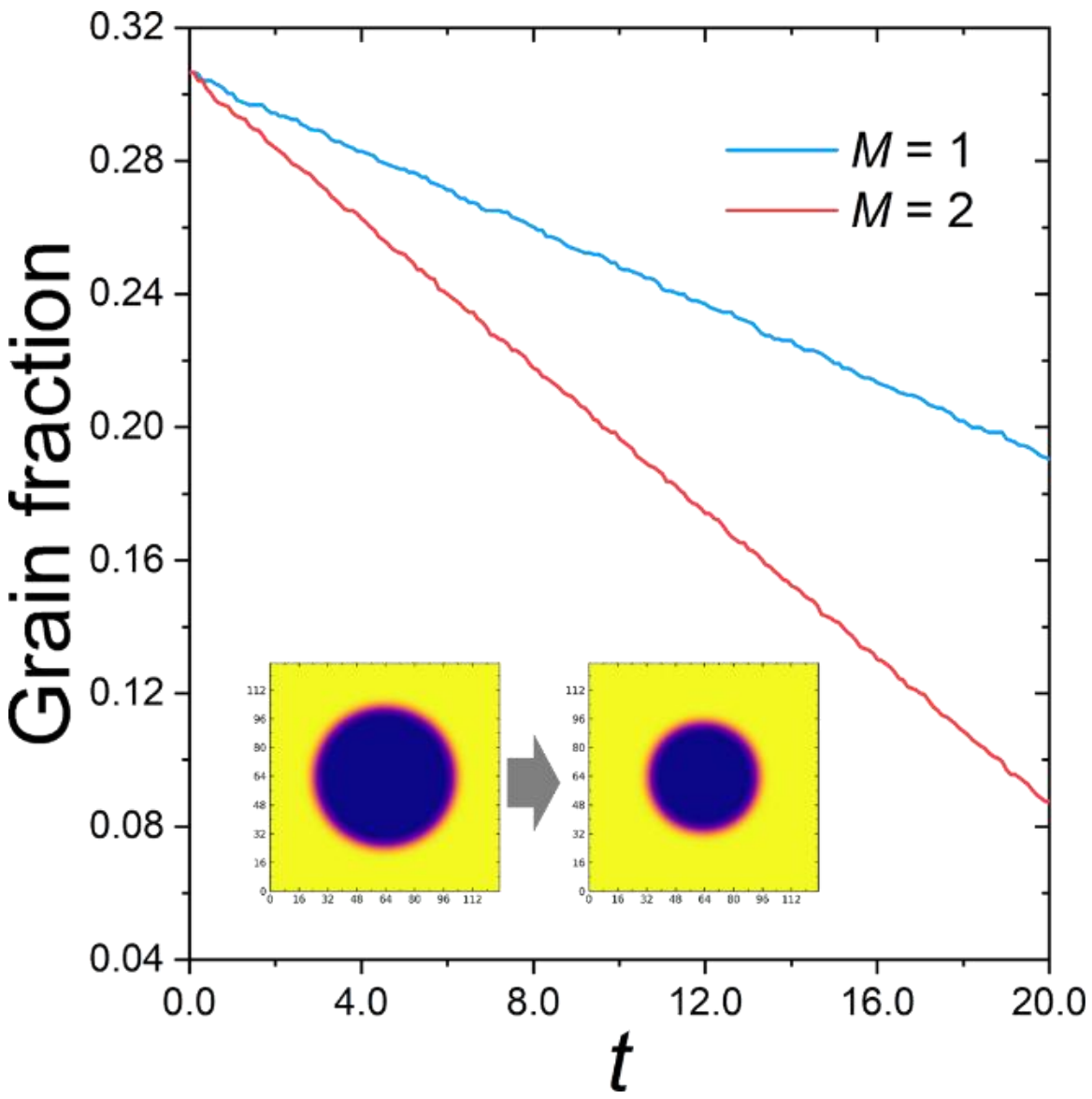

**FIG 11.** Time evolution of the grain fraction of a shrinking circular grain.

### D. Wulff Constructions

The grain shape is primarily controlled by the anisotropic GB function $B_{aniso}(\theta^+, \theta^-, \boldsymbol{v})$. In this section, we investigate the Wulff shapes of an initially circular grain for different values of the anisotropic coefficient $\epsilon_4$. The simulation setup is similar to that used in Section IV C, and four-fold inclination symmetry is imposed. The orientations of the two grains are chosen such that $\frac{\theta^+ + \theta^-}{2} = 0$, ensuring that the simulations isolate inclination effects in the reference frame only. The resulting anisotropic grain shapes are compared to and measured by the analytical Wulff shapes constructed in Appendix C.

During the simulations, the GB migrates. Notably, the inclination equilibrates at an early stage of the migration process (i.e., $t \approx 5$), ensuring that the shrinking grain maintains a self-similar shape during subsequent evolution. As a result, the measured anisotropy is not affected by GB migration. **FIG 12** compares the simulated grain shapes with the analytical Wulff shapes for different values of the anisotropic coefficient. As $\epsilon_4$ increases beyond 0.6, nonphysical 'ears' appear in the analytical Wulff shapes, simply indicating that certain interface orientations are missing from the construction. In Appendix C, we show that the analytical Wulff construction loses convexity when $E_4 > \frac{1}{15}$ (i.e., $\epsilon_4 > 0.6$), where $E_4$ is the anisotropic coefficient from the analytical Wulff construction. This critical value has also been reported in previous studies [6] and the equilibrium grain shape remains convex in all cases, as it should.

As observed in **FIG 12**, the grains are rotated by a certain angle rather than being symmetric with respect to the x- and y-axes. Approximating the grain shape as a square, the GB inclinations of its four facets can be measured as $\Psi_i \approx -0.4 + i\frac{\pi}{2}$, where index $i = 0,1,2,3$. This rotation arises because the GB energy is minimized at this set of inclination angles. For a four-fold symmetric grain, the GB energy is approximated using Eq. (4.2) as:

$$F_4 = 2\sum_{i=0}^{1} F_0(a+b)|\sin(n\Delta\theta)| \left[1 + \frac{b}{a+b}\epsilon_4 \sin(4\Psi_i)\right], \quad (4.4)$$

where the misorientation $\Delta\theta$ is fixed and $F_0$ is a constant proportional to the GB length. The summation is carried out over $i = 0$ to $1$ and multiplied by a factor of 2 to account for the symmetry of the grain. It follows that $F_4$ has a periodicity of $\frac{\pi}{2}$ and attains its minimum values at $\Psi_0 \approx -0.4$ and $\Psi_1 \approx -0.4 + \frac{\pi}{2}$, in excellent agreement with the simulations shown in **FIG 12**. Similar behavior is expected for grains with higher symmetry, although these cases are not shown in this paper.

**FIG 13** illustrates the relationship between the anisotropic coefficient $\epsilon_4$ in the proposed model and the corresponding analytical coefficient $E_4$ for different misorientations. It shows that smaller misorientation leads to stronger anisotropy, which is consistent with the form of $B_{aniso}(\theta^+, \theta^-, \boldsymbol{v})$: as the misorientation decreases, the relative contribution of the inclination term increases.

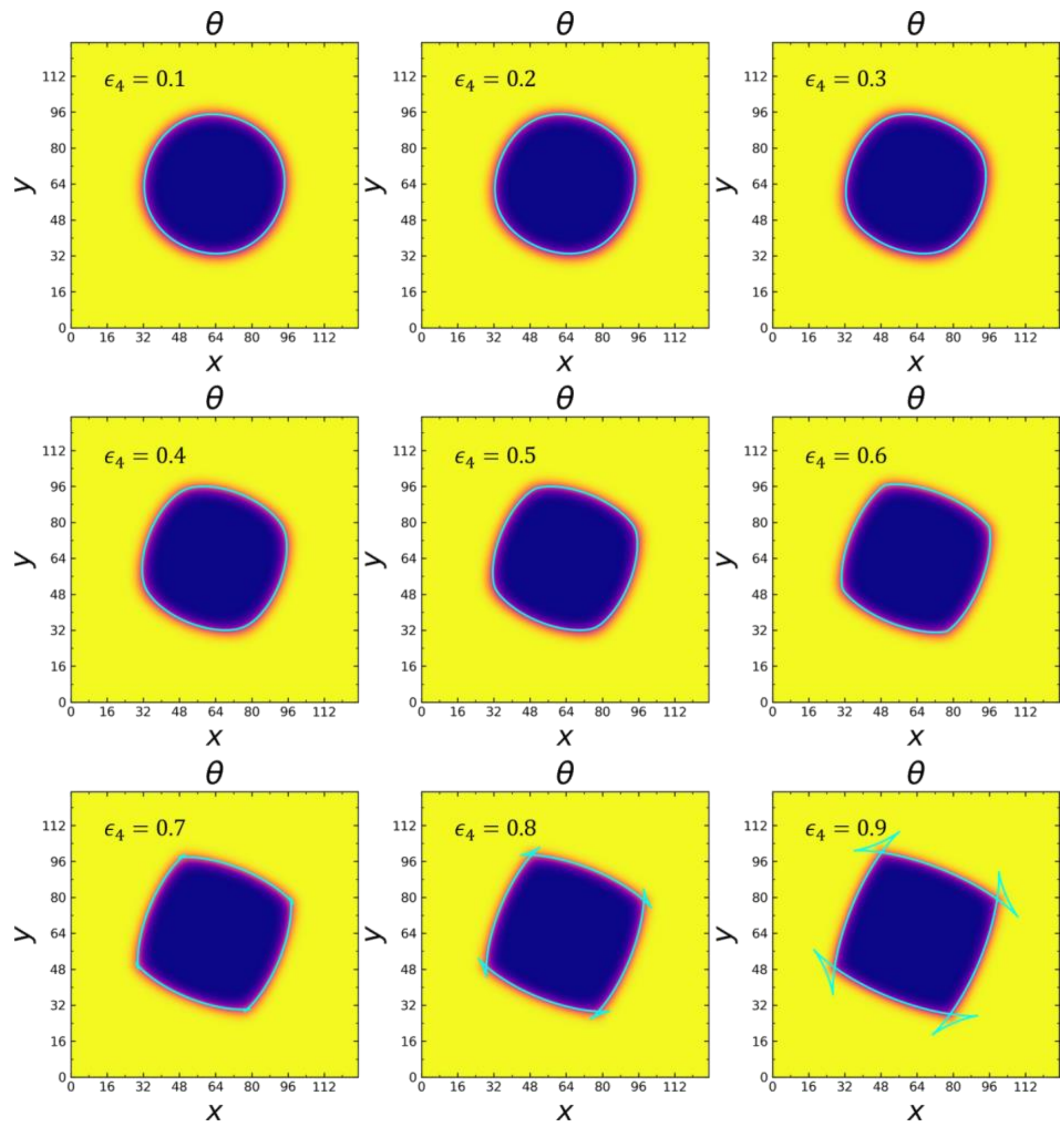

**FIG 12.** Wulff constructions (cyan curves) for anisotropic grains with misorientation $\Delta\theta = 0.5$ and varying anisotropic coefficient $\epsilon_4$. Note that the actually Wulff shapes do not include the 'ears'.

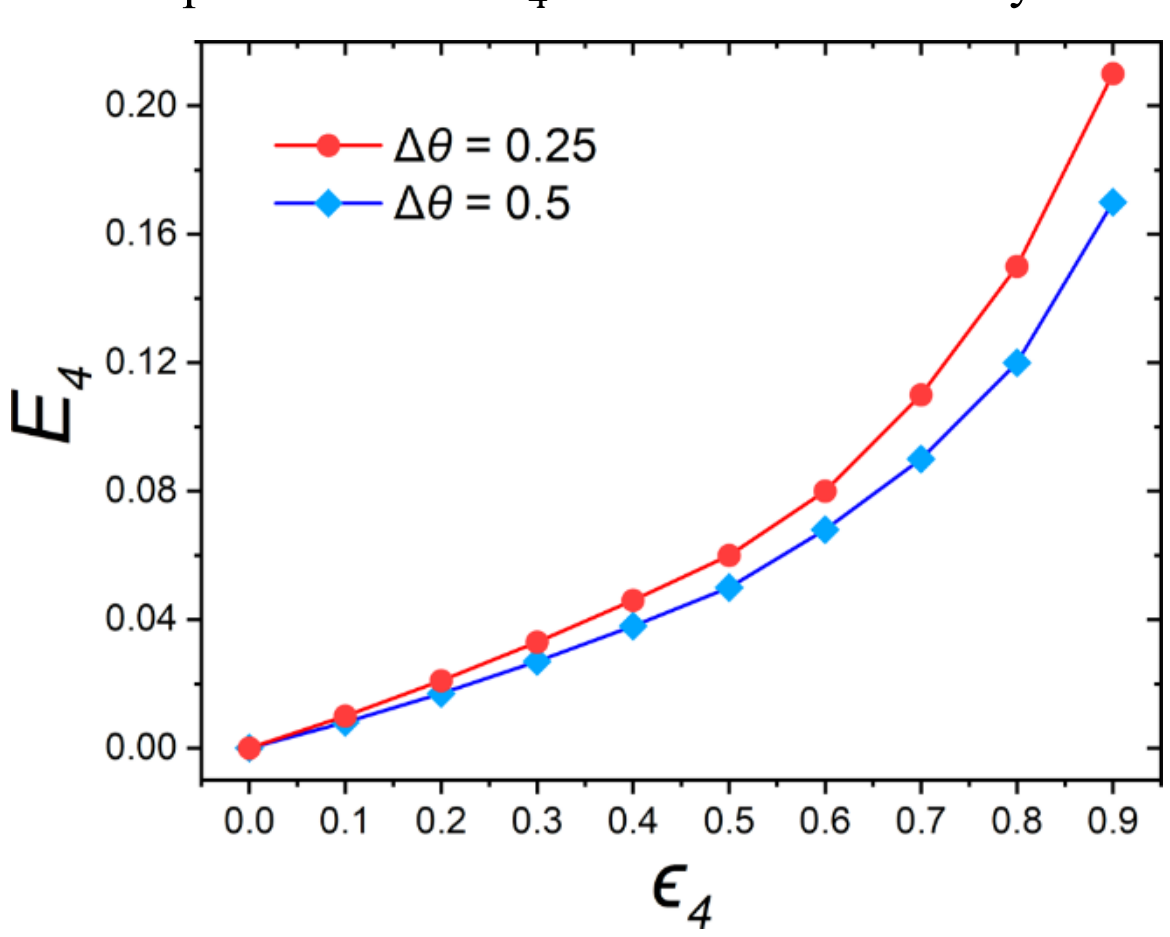

**FIG 13.** Relationship between the anisotropic coefficient $\epsilon_4$ in the anisotropic GB function $B_{aniso}$ and the corresponding coefficient $E_4$ in the classical GB energy Eq. (C.1).

**E. Triple Junctions**

A triple junction is the set of points where three grains meet. Simulations involving triple junctions provide further validation of the influence of the GB functions on more complex grain shapes. Accurately modeling the properties of a triple junction is also essential for extending the proposed model to general polycrystalline systems. In this section, we examine the dihedral angles obtained from the simulations. The theoretical dihedral angles are computed using the Herring force balance [40] at a triple junction:

$$\sum_{i=1}^{3} \gamma_i \boldsymbol{t_i} + \frac{\partial \gamma_i}{\partial \Psi_i} \boldsymbol{v_i} = \boldsymbol{0}, \tag{4.5}$$

where index $i = 1,2,3$ denotes the three interfaces meeting at the triple junction. $\gamma_i$ is the GB energy density, $\boldsymbol{t_i}$ is the unit tangent vector, and $\boldsymbol{v_i}$ is the unit normal vector of the $i$-th GB. We begin with isotropic models, where the GB energy density is given by:

$$\gamma_i = (a + b)|\sin(n\Delta\theta_i)|, \tag{4.6}$$

where the constants $a$ and $b$ are specified in Section IV B, and $\Delta\theta_i$ denotes the misorientation angle of the $i$th GB. Because inclination dependence is absent in Eq. (4.6), the derivative term $\frac{\partial \gamma_i}{\partial \Psi_i}$ in Eq. (4.5) vanishes. Consequently, for isotropic models, the Herring force balance reduces to:

$$\sum_{i=1}^{3} \gamma_i \boldsymbol{t_i} = \boldsymbol{0}. \tag{4.7}$$

Using the law of sines, Eq. (4.7) can be transformed into Young's law [40]:

$$\frac{\gamma_1}{\sin \varphi_1} = \frac{\gamma_2}{\sin \varphi_2} = \frac{\gamma_3}{\sin \varphi_3}. \tag{4.8}$$

where $\varphi_1, \varphi_2, \varphi_3$ are the dihedral angles between tangent vectors $\boldsymbol{t_2}$ and $\boldsymbol{t_3}$, $\boldsymbol{t_1}$ and $\boldsymbol{t_3}$, $\boldsymbol{t_1}$ and $\boldsymbol{t_2}$, respectively. Since $\gamma_i$ has a simple sinusoidal dependence on misorientation, Eq. (4.8) can be solved straightforwardly.

The simulations are performed using the triplet system shown in **FIG 14**. The grains are labeled such that their lattice orientations vary monotonically, either clockwise or counterclockwise, from the purple region to the yellow region. Note that grain #1 (purple) and grain #3 (yellow) are relatively secondary to grain #2 (pink), and therefore tend to shrink during the GB migration. Importantly, no GBs are artificially fixed in the simulations. As discussed in Section IV D, the GB inclination equilibrates at an early stage of the evolution. Consequently, both the inclination and the dihedral angles remain unchanged during the subsequent migration process.

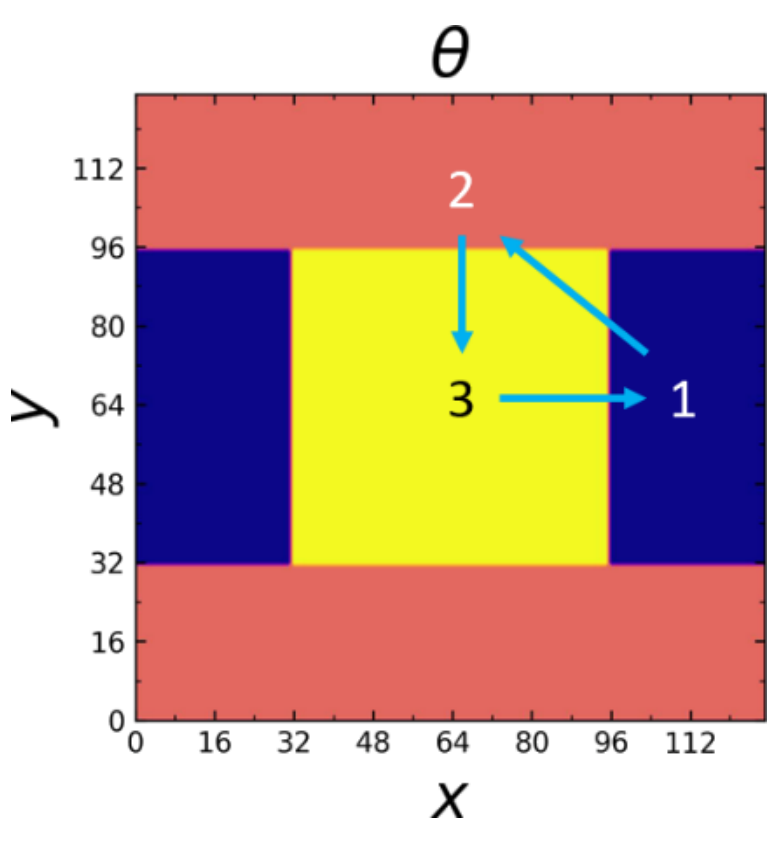

**FIG 14.** The triplet system used in the triple junction simulations. The GB tangents between grain #1 and #2, #2 and #3, and #3 and #1 are denoted by $\boldsymbol{t_1}$, $\boldsymbol{t_2}$, and $\boldsymbol{t_3}$, respectively. The corresponding misorientations are represented as $\Delta\theta_1$, $\Delta\theta_2$, and $\Delta\theta_3$. The dihedral angle between any two GB tangents $\boldsymbol{t_i}$ and $\boldsymbol{t_j}$ is denoted by $\varphi_k$, where $i \neq j \neq k$.

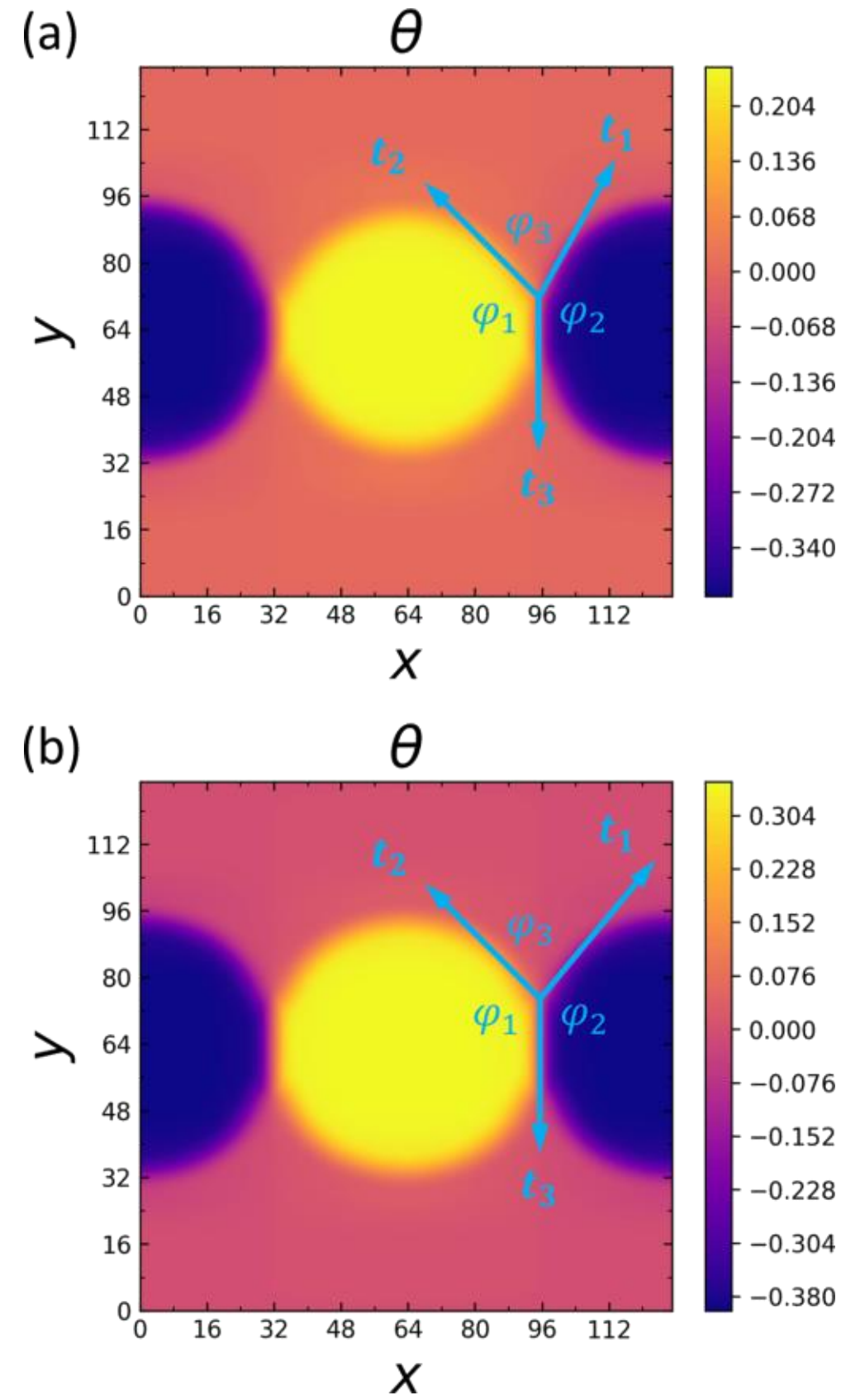

**FIG 15.** Measurements for the dihedral angles at triple junctions with misorientations (a) $\Delta\theta_1 = 0.4$, $\Delta\theta_2 = 0.25$, $\Delta\theta_3 = 0.65$ and (b) $\Delta\theta_1 = 0.4$, $\Delta\theta_2 = 0.35$, $\Delta\theta_3 = 0.75$. The frames are taken at the same simulation time. Blue arrows represent the GB tangents.

Simulations are carried out for five different triplet configurations. **FIG 15** presents the simulated dihedral angles for two representative cases, while **TABLE I** summarizes the results for all five triplets, comparing the simulated dihedral angles with their corresponding theoretical values. Excellent agreement between the simulated and theoretical values is observed in all cases.

As shown in **TABLE I**, dihedral angles associated with lower-angle GBs are larger than those corresponding to higher-angle GBs. Taking $\varphi_2$ as an example, comparison of **FIGs 15(a)** and **15(b)** shows that $\varphi_2$ in **FIG 15(a)** is larger than in **FIG 15(b)**. This difference is because the misorientation $\Delta\theta_2$ between grains #3 and #2 is smaller in **FIG 15(a)**, whereas the misorientation $\Delta\theta_1$ between grains #1 and #2 is identical in both configurations. Consequently, the radius of grain #3 is smaller in **FIG 15(a)**, while the radius of grain #1 remains unchanged. The resulting difference in the radius of grain #3 leads to the observed variation in $\varphi_2$, whereas the unchanged radius of grain #1 results in identical values of $\varphi_1$ in the two cases.

For anisotropic models, because the inclination dependence is included, the theoretical dihedral angles must be computed using the full Herring force balance Eq. (4.5). The Herring force balance can be decomposed into components along the $x$ and $y$ directions as follows:

$$\begin{cases} \sum_{i=1}^{3} -\gamma_i \sin \Psi_i + \frac{\partial \gamma_i}{\partial \Psi_i} \cos \Psi_i = 0 \\ \sum_{i=1}^{3} \gamma_i \cos \Psi_i + \frac{\partial \gamma_i}{\partial \Psi_i} \sin \Psi_i = 0 \end{cases}, \quad (4.9)$$

where $\Psi_i$ is the inclination angle between the normal of the $i$th GB and the x-axis. Once the inclination angles are obtained, the three dihedral angles can be given by:

$$\begin{aligned} \varphi_1 &= \Psi_3 - \Psi_2 \\ \varphi_2 &= 2\pi - \Psi_3 + \Psi_1, \\ \varphi_3 &= \Psi_2 - \Psi_1 \end{aligned} \quad (4.10)$$

**TABLE II** summarizes the simulated and theoretical dihedral angles for anisotropic models with $\epsilon_4 = 0.5$ . In contrast to the isotropic cases, the dihedral angles $\varphi_1$ and $\varphi_2$ differ significantly even when the corresponding misorientations $\Delta\theta_1$ and $\Delta\theta_2$ are equal. Notably, owing to the lack of symmetry of the dihedral angles with respect to misorientation, multiple dihedral angle solutions exist for a given set of misorientations. Each solution corresponds to a distinct triple junction geometry. This multiplicity of solutions is also expected from a mathematical perspective, as Eq. (4.9) has multiple solution sets. **TABLE II** reports only the dihedral angles corresponding to the same triple junctions listed in **TABLE I**. Overall, the simulation results remain in excellent agreement with the theoretical values.

**TABLE I.** Comparison between simulated and theoretical (in parentheses) dihedral angles for isotropic triple junctions.

| $\Delta\theta_1$ (rad) | $\Delta\theta_2$ (rad) | $\Delta\theta_3$ (rad) | $\varphi_1{}^\circ$ | $\varphi_2{}^\circ$ | $\varphi_3{}^\circ$ |
|---|---|---|---|---|---|
| 0.25 | 0.25 | 0.50 | 150.95° (151.35°) | 150.95° (151.35°) | 58.11° (57.30°) |
| 0.35 | 0.35 | 0.70 | 140.19° (139.89°) | 140.19° (139.89°) | 79.62° (80.22°) |
| 0.35 | 0.25 | 0.60 | 140.19° (139.89°) | 150.95° (151.35°) | 68.86° (68.76°) |
| 0.40 | 0.25 | 0.65 | 135.00° (134.16°) | 150.95° (151.35°) | 74.05° (74.49°) |
| 0.40 | 0.35 | 0.75 | 135.00° (134.16°) | 140.19° (139.89°) | 84.81° (85.95°) |

**TABLE II.** Comparison between simulated and theoretical (in parentheses) dihedral angles for anisotropic triple junctions with $\epsilon_4 = 0.5$.

| $\Delta\theta_1$ (rad) | $\Delta\theta_2$ (rad) | $\Delta\theta_3$ (rad) | $\varphi_1{}^\circ$ | $\varphi_2{}^\circ$ | $\varphi_3{}^\circ$ |
|---|---|---|---|---|---|
| 0.25 | 0.25 | 0.50 | 122.90° (123.68°) | 166.93° (165.72°) | 70.17° (70.60°) |
| 0.35 | 0.35 | 0.70 | 115.90° (115.61°) | 159.55° (159.50°) | 84.55° (84.89°) |
| 0.35 | 0.25 | 0.60 | 115.90° (115.55°) | 164.07° (165.17°) | 80.02° (79.28°) |
| 0.40 | 0.25 | 0.65 | 113.93° (112.36°) | 164.07° (165.15°) | 81.99° (82.50°) |
| 0.40 | 0.35 | 0.75 | 113.93° (112.43°) | 159.55° (159.46°) | 86.51° (88.11°) |

### F. Polycrystalline Simulations

In this section, we examine this model's behavior in more complex systems. A polycrystalline system composed of six grains is constructed, as illustrated in **FIG 16**. The system is generated using a Voronoi tessellation based on randomly distributed seed points. All grain orientations are assigned within the range $\theta \in [-0.6, 0.18]$ to avoid topological defects [24]. This simulation setup is sufficiently general to qualitatively reproduce grain growth, coalescence, and triple junction behavior.

Two simulations are performed for isotropic and anisotropic systems, respectively, with the anisotropic coefficient set to $\epsilon_4 = 0.5$ . Selected frames from both simulations are shown in **FIG 17**. The evolution illustrates the growth of initially larger grains and the coalescence of smaller ones. Grains #1 and #4 are relatively small compared to the others and therefore tend to shrink and eventually merge with their neighboring grains. **FIGs 17(a-b)** and **17(b-c)** capture the coarsening processes of grains #1 and #4, respectively. Moreover, due to the lower energy barrier, small grains are more likely to merge with neighboring grains that share lower-angle GBs.

**FIG 18** plots the time evolution of the area fractions of the six grains in the isotropic system. Grain #1 completely disappears at approximately $t \approx 8$ (indicated by black arrow), leading to an increase in the area fractions of its two neighboring grains (#2 and #3), which share lower-angle GBs with grain #1. These increases are indicated by the protuberance marked with blue arrows. A similar coarsening behavior is observed for grain #4. Since grains #3 and #5 are its neighboring grains with lower-angle GBs, the disappearance of grain #4 (marked by red arrows) results in noticeable increases in the area fractions of grains #3 and #5 at approximately $t \approx 17$, as indicated by green arrows.

The dependence of dihedral angles on misorientation is also evident in the polycrystalline simulations. The arrow triplets in **FIGs 17(b)** and **17(c)** indicate the GB tangents at two selected triple junctions. In each triplet, the red arrow represents the GB tangent corresponding to the largest misorientation ($\Delta\theta_3$) among the three GBs. In both cases, this GB is associated with the smallest dihedral angle. Specifically, tangent triplet (i) corresponds to a triple junction where $\Delta\theta_1 = \Delta\theta_2$, and therefore the dihedral angles corresponding to the two blue tangents are nearly equal. In contrast, triplet (ii) corresponds to a triple junction where $\Delta\theta_1 \neq \Delta\theta_2$, resulting in clearly different dihedral angles. These observations are consistent with the results summarized in **TABLE I**.

The presence of anisotropy leads to significant changes in grain shape. For GBs, the black arrows in **FIGs 17(b)** and **17(e)** indicate that, in the anisotropic system, the curvature of low-angle GBs changes significantly, whereas the white arrows indicate that higher-angle GBs are less sensitive to anisotropic effects. As a result, grain shapes in the anisotropic system are more square-like, particularly for grains bounded by low-angle GBs. This observation is consistent with the trend shown in **FIG 13**. For triple junctions, as indicated by the green arrows in **FIGs 17(c)** and **17(f)**, the low-angle GB between the purple and pink regions in the anisotropic system is nearly perpendicular to the other two GBs. Compared to the isotropic system, the anisotropic dihedral angles corresponding to the two higher-angle GBs are closer to each other. This observation is also consistent with the trends shown in **TABLES I** and **II**, where the difference between $\varphi_3$ and $\varphi_1$ in **TABLE II** is much smaller than in **TABLE I**.

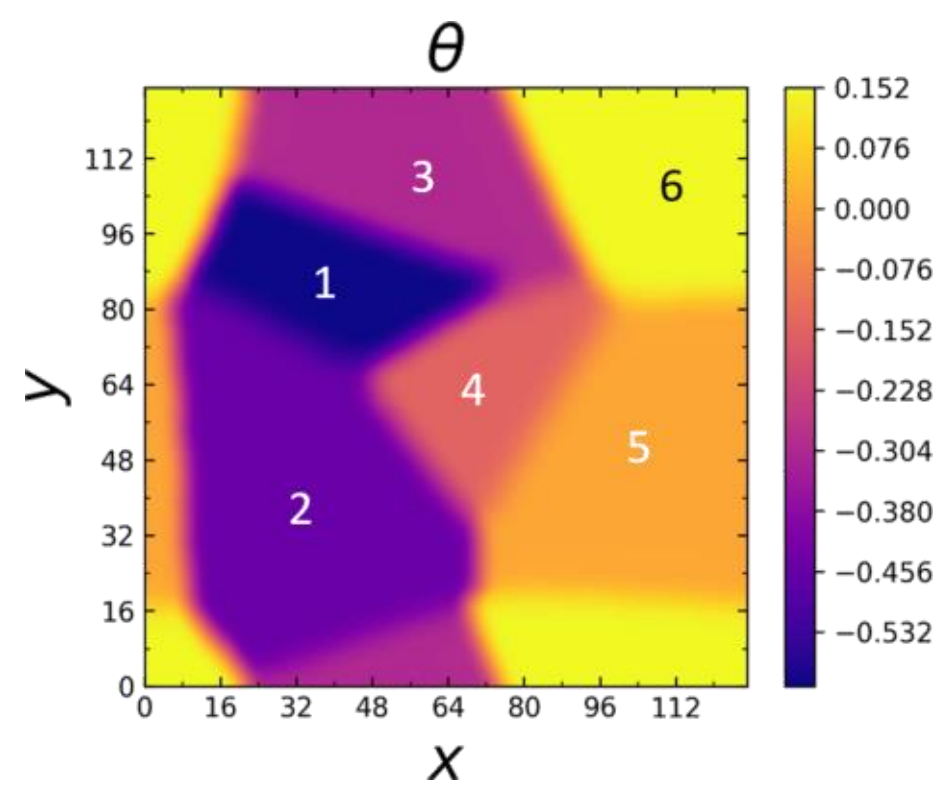

**FIG 16.** System used for polycrystalline simulations. Grains are numbered according to their orientation values, ranked from lowest to highest.

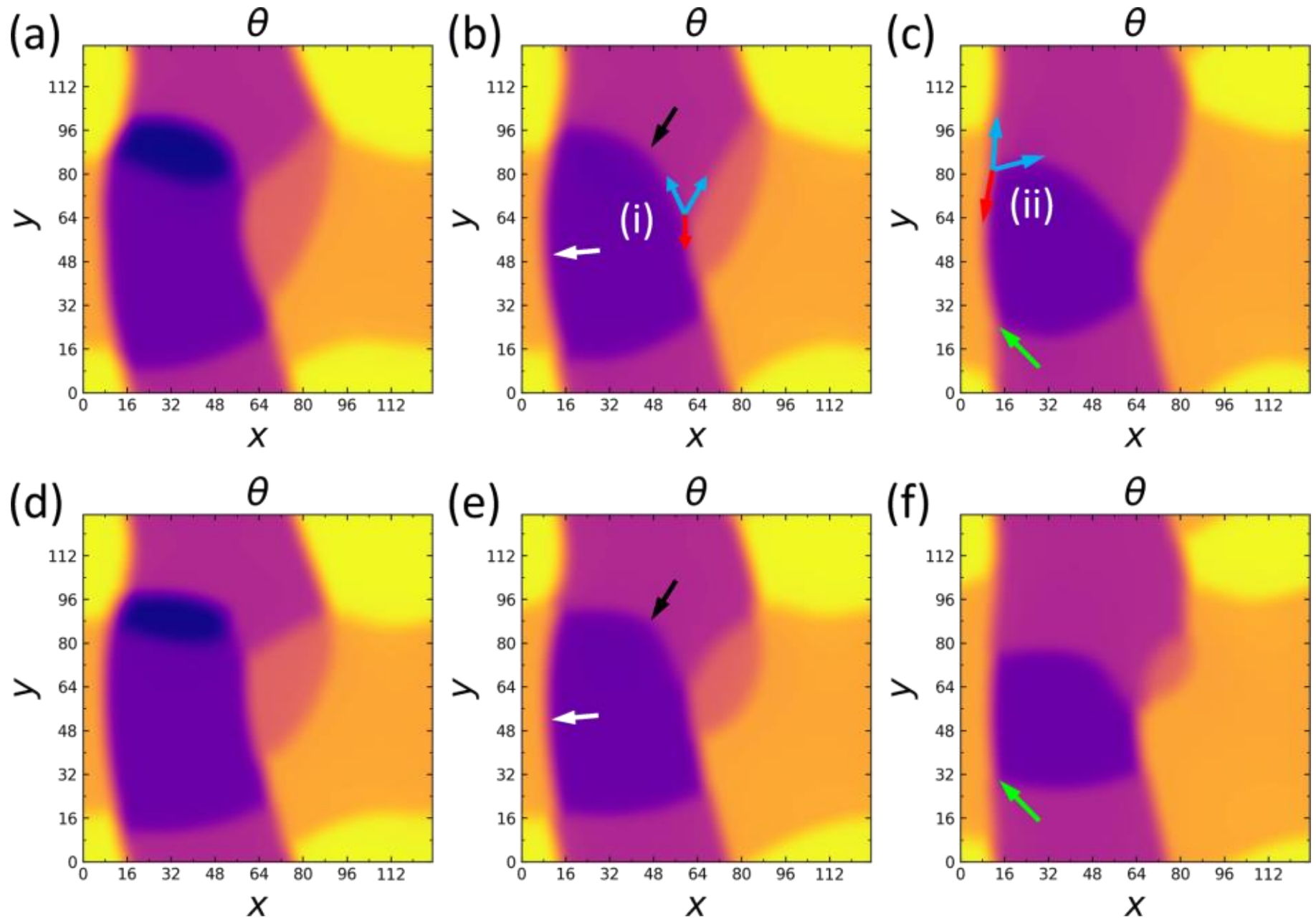

**FIG 17.** Selected frames from the polycrystalline simulations. The first row (a-c) shows results from the isotropic system, while the second row (d-f) shows the anisotropic system with $\epsilon_4 = 0.5$. Each column corresponds to the same simulation time for both systems.

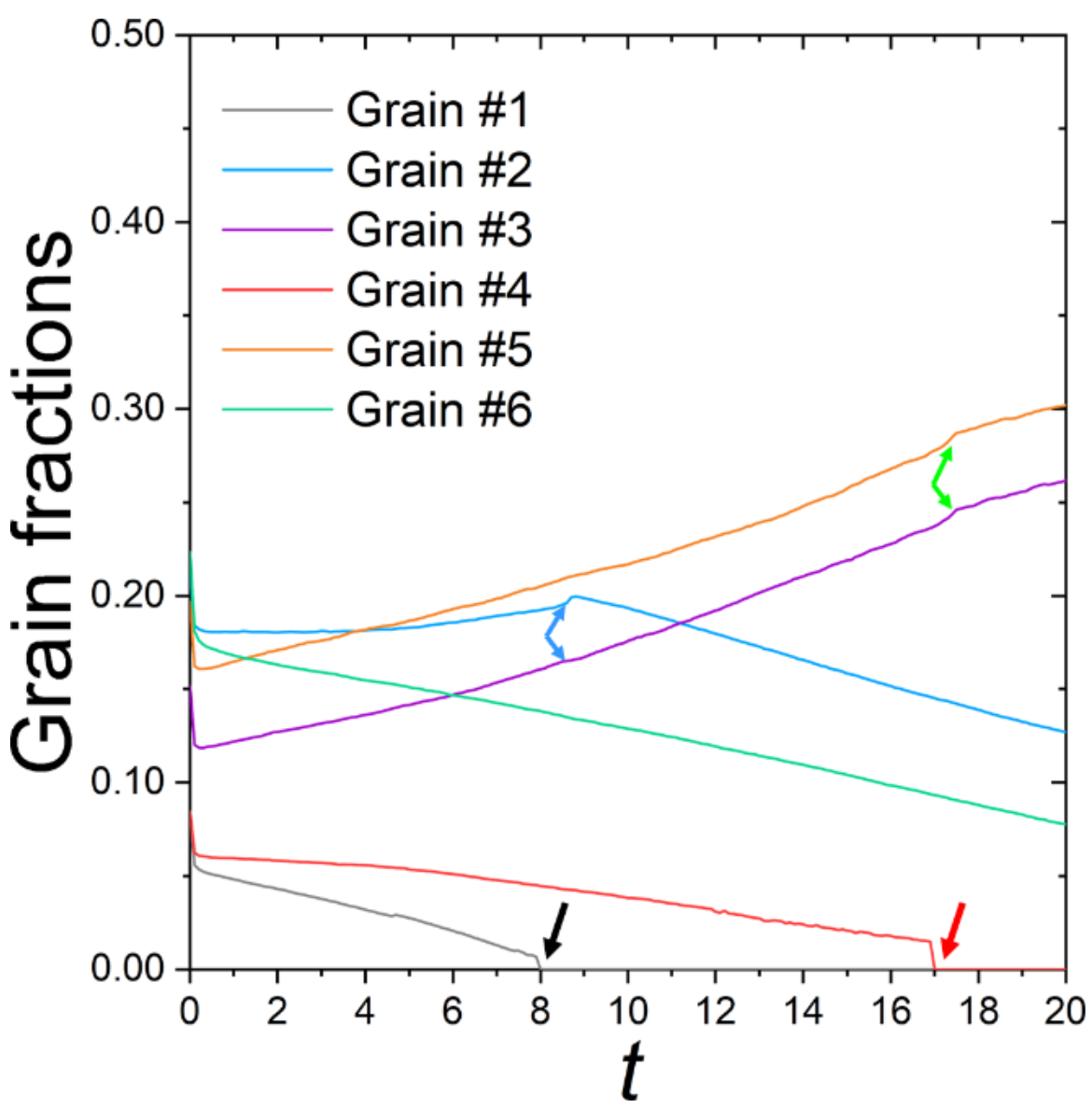

**FIG 18.** Time evolution of the area fractions of the six grains in the isotropic simulation system.

## V. EXTENSION TO 3D

This model can also be extended to three dimensions. In 3D systems, the orientation field must represent rotations in $\mathbb{R}^3$ and can no longer be described by a scalar. Instead, we use the quaternion [41-43] $\boldsymbol{q} \in \mathbb{R}^{1\times 4}$ to represent the orientation field:

$$\boldsymbol{q} = \left[\cos\frac{\theta}{2} \quad \boldsymbol{u}^T \sin\frac{\theta}{2}\right], \tag{5.1}$$

where $\theta$ is the rotation angle and $\boldsymbol{u} \in \mathbb{R}^3$ is a unit vector denoting the rotation axis. Compared with alternative representations such as Euler angles and rotation matrices, quaternions require only four variables to store rotational information, whereas rotation matrices ($3 \times 3$ matrices) require nine. Moreover, quaternions avoid gimbal lock [44], a well-known limitation of Euler angle representations. The quaternion defined in Eq. (5.1) satisfies the unit-norm constraint $|\boldsymbol{q}| = 1$, which allows the number of independent variables to be reduced from four to three. Consequently, only the last three components (the 'implicit' part) of the quaternion need to be tracked. Writing $\boldsymbol{q} = [q_1 \quad q_2 \quad q_3 \quad q_4]$, the first component $q_1$ can be expressed in terms of the remaining components as:

$$q_1 = \sqrt{1 - \sum_{i=2}^{4} {q_i}^2}. \tag{5.2}$$

The gradient of $q_1$ then follows as:

$$\nabla q_1 = -\frac{\sum_{i=2}^{4} q_i \nabla q_i}{\sqrt{1 - \sum_{i=2}^{4} {q_i}^2}}. \tag{5.3}$$

The free energy of GBs retains the same structural form as in the two-dimensional case but becomes a functional of quaternions:

$$F = \int \mathrm{d}\boldsymbol{x} \left[ B_{aniso}(\boldsymbol{q}^+, \boldsymbol{q}^-, \boldsymbol{v}) w(|\phi|) \overline{|\phi|}^2 + B_{iso}(\boldsymbol{q}^+, \boldsymbol{q}^-)(1 - w(|\phi|)) \left( \alpha \overline{|\phi|}^2 + \beta c(\boldsymbol{q}, \boldsymbol{q}^+, \boldsymbol{q}^-) \right) \right], \tag{5.4}$$

where $\phi \in \mathbb{R}^3$ is a vector characterizing the orientation of the GB. It is defined as $\phi = \nabla \boldsymbol{q} \boldsymbol{g}^T$, with $\nabla \boldsymbol{q} \in \mathbb{R}^{3\times4}$ and $\boldsymbol{g} \in \mathbb{R}^{1\times4}$ being the unit quaternion that maximizes $\boldsymbol{g}(\nabla \boldsymbol{q}^T \nabla \boldsymbol{q})\boldsymbol{g}^T$. The quaternion $\boldsymbol{g}$ corresponds to the eigenvector associated with the largest eigenvalue of the matrix $\nabla \boldsymbol{q}^T \nabla \boldsymbol{q}$. In the 3D formulation, the misorientation between two adjoining grains is defined as $2\arccos(\boldsymbol{q}^+ \cdot \boldsymbol{q}^-)$, where $\boldsymbol{q}^\pm$ are the nonlocal quaternion fields extracted from the bulk lattice orientations of the adjoining grains.. The GB normal is given by $\boldsymbol{v} = \frac{\phi}{|\phi|}$. The double-well potential becomes a function of the quaternion fields and simultaneously serves as a constraint to enforce the unit-norm condition $|\boldsymbol{q}| = 1$. The GB functions $B_{aniso}(\boldsymbol{q}^+, \boldsymbol{q}^-, \boldsymbol{v})$ and $B_{iso}(\boldsymbol{q}^+, \boldsymbol{q}^-)$ need to be further developed to incorporate 3D crystallographic symmetries, which are significantly more complex than their 2D counterparts. The implementation and verification of the full 3D model will be addressed in future work.

## VI. DISCUSSION

Notably, by incorporating orientation information from nonlocal points, the total GB energy can be formulated solely in terms of the orientation field $\theta$, which substantially reduces memory usage. Moreover, a wide range of forms of GB energy can be realized by selecting proper GB functions with the corresponding form. In addition, the proposed model is shown to be frame indifferent, and the asymptotic analysis provides analytical justification for the model.

This model can also be implemented within the finite element method (FEM) framework. The overall implementation workflow is illustrated in **FIG 19**. The only nonstandard step is the computation of the nonlocal orientation fields $\theta^\pm$ prior to assembling the residual. Nevertheless, this step is feasible, as demonstrated by our successful simulations of a anisotropic two-grain system. Further development and validation of the FEM implementation will be presented in our next paper.

Beyond the benchmark cases we consider herein, the proposed model is suited for modeling equilibrium grain morphologies and microstructure evolution in real alloys, such as refractory hexagonal close-packed alloys [45-47]. The framework also provides a natural pathway for coupling grain boundary energetics with temperature and stress fields, enabling studies of microstructure evolution under thermal annealing and mechanical loading. In addition, the model offers a direct bridge between atomistic and mesoscale descriptions, as GB functions can be parameterized using GB energy data obtained from atomistic calculations, including empirical and machine-learning-based interatomic potentials [48-50].

This model also has certain limitations. Compared with existing models, it has a more complex formulation and involves highly nonlinear terms. As a result, very small time steps are required for numerical stability (as reflected by the ultra-small Courant number in Section III E), leading to increased computational cost. In addition, the nonlocal orientation fields $\theta^+$ and $\theta^-$ have to be computed at certain time intervals. To address these computational challenges, we have developed a parallel C++ implementation that enables simulations to be completed within a reasonable time frame. Further performance improvements may be achieved by integrating the model into established open-source phase-field frameworks such as Moose [51-52], OpenPhase [53-54], MMSP [55]. The FEM formulation of this model provides a solid foundation for such integrations, which will be our future work.

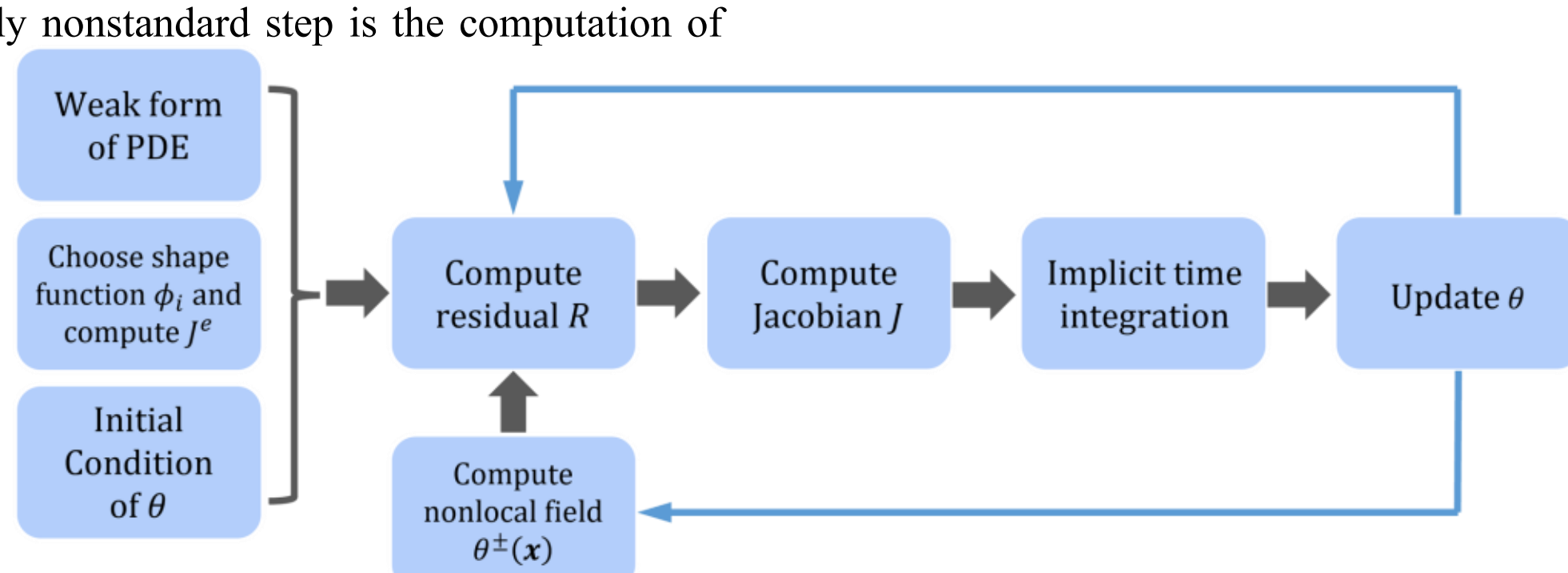

**FIG 19.** Workflow for the finite element method (FEM) implementation.

## VII. CONCLUSIONS

In summary, we have developed an orientation field phase-field model that incorporates grain boundary anisotropy through a nonlocal functional of a single orientation field. Explicit GB functions are introduced to provides direct control of the misorientation- and inclination-dependence of the GB energy. The model accurately reproduces key benchmark behaviors, including linear grain growth kinetics, equilibrium Wulff shapes with prescribed anisotropy, and analytical dihedral angles at triple junctions. Simulations of polycrystalline systems further qualitatively reproduce grain growth and coalescence, capture distinct Wulff shapes associated with low-angle and high-angle GBs, and reflect the relationship between dihedral angles and grain misorientations at triple junctions. A formulation for extending this model to 3D is also presented at the end.

## ACKNOWLEDGMENTS

This research was funded by the US Army Research Office under grant number W911NF-21-2-0161. Computational resources were provided by the Center for Computation and Visualization at Brown University.

## APPENDIX A

The steady-state solution is derived for 1D systems. We define the steady-state free energy:

$$F^* = \int \mathrm{d}x \times \left( S|\nabla\theta|^2 + C\left(\frac{\theta-\theta^-}{\theta^+-\theta^-}\right)^2 \left[1-\left(\frac{\theta-\theta^-}{\theta^+-\theta^-}\right)\right]^2 \right), \tag{A.1}$$

where all auxiliary functions are taken to be constants. Parameters $C$ and $S$ are fitting constants that control the GB width. The steady state can be found by solving $\delta F^* = 0$:

$$\frac{\delta F^*}{\delta\theta} = \frac{\partial f^*}{\partial\theta} - \nabla\cdot\frac{\partial f^*}{\partial(\nabla\theta)} = 0. \tag{A.2}$$

This gives:

$$2C\frac{(\theta-\theta^+)(\theta-\theta^-)(2\theta-\theta^+-\theta^-)}{(\theta^+-\theta^-)^4} - 2S\nabla^2\theta = 0. \tag{A.3}$$

Assuming that the solution has the form:

$$\theta(x) = A + \mathcal{B}\tanh[\lambda(x-x_0)], \tag{A.4}$$

we can then obtain:

$$\nabla^2\theta = -2\mathcal{B}\lambda^2\tanh[\lambda(x-x_0)] \times (1-\tanh^2[\lambda(x-x_0)]), \tag{A.5}$$

where $x_0$ is determined by the GB position and $\lambda$ is the reciprocal of the equilibrium GB width. Substituting Eq. (A.4) and Eq. (A.5) into Eq. (A.3), we can solve the constants in $\theta$:

$$A = \frac{\theta^+ + \theta^-}{2},\quad \mathcal{B} = \frac{\theta^+ - \theta^-}{2},\quad \lambda = \frac{1}{2(\theta^+ - \theta^-)}\frac{\sqrt{C}}{\sqrt{S}}. \tag{A.6}$$

Nonlocal orientation fields $\theta^\pm$ can also be extrapolated in '$\theta$-space'. We define:

$$\xi \coloneqq \tanh[\lambda(x-x_0)],$$

$$e \coloneqq \theta - A,\qquad g \coloneqq \nabla\theta,\qquad \hbar \coloneqq \nabla^2\theta. \tag{A.7}$$

Thus, we have the following relationships:

$$e = \mathcal{B}\xi, \tag{A.8}$$

$$g = \mathcal{B}\lambda(1-\xi^2), \tag{A.9}$$

$$\hbar = -2\mathcal{B}\lambda^2\xi(1-\xi^2). \tag{A.10}$$

Combining (A.8), (A.9), and (A.10), we can solve for $e$, $\mathcal{B}$, and $\xi$:

$$e = -\frac{2g^2\hbar}{(2g\lambda+\hbar)(2g\lambda-\hbar)}, \tag{A.11}$$

$$\mathcal{B} = \frac{4g^3\lambda}{(2g\lambda+\hbar)(2g\lambda-\hbar)}, \tag{A.12}$$

$$\xi = -\frac{\hbar}{2g\lambda}. \tag{A.13}$$

Moreover, since $\theta^\pm = A \pm \mathcal{B}$, we obtain:

$$\theta^+ = \theta + \frac{2|\nabla\theta|^2}{2\lambda\nabla\theta - \nabla^2\theta}, \tag{A.14}$$

$$\theta^- = \theta - \frac{2|\nabla\theta|^2}{2\lambda\nabla\theta + \nabla^2\theta}. \tag{A.15}$$

Note that when the point $x$ is in the far field (i.e., $|\nabla\theta| \to 0$), we have $\theta^\pm(x) \to \theta(x)$.

## APPENDIX B

Following a similar procedure to that in REF [6], the solution domain of Eq. (3.1) can be divided into two regions: the GB and the bulk regions. In the bulk region, the solution for $\theta$ is trivial, which is simply $\theta = \theta^+$ or $\theta = \theta^-$. In the GB region, the PDE is reformulated in a curvilinear coordinate system defined with respect to a fixed laboratory frame. In this coordinate system, $r$ denotes the coordinate normal to the interface, while $s$ represents the arclength tangential to the interface, with $r = 0$ corresponds to the interface itself.

Under the laboratory frame, the time derivative appearing in Eq. (3.1) is expressed as:

$$\left.\frac{\partial\theta}{\partial t}\right|_{lab} = \left.\frac{\partial\theta}{\partial t}\right|_{r,s} - V_n\frac{\partial\theta}{\partial r}, \tag{B.1}$$

where $\left.\frac{\partial\theta}{\partial t}\right|_{r,s}$ denotes the time derivative in the local material frame, and $V_n$ is the velocity of the interface along the normal direction. In the GB region, the normal velocity $V_n$ can be assumed to be constant. Substituting Eq. (B.1) into Eq. (3.1), we have:

$$\left.\frac{\partial\theta}{\partial t}\right|_{r,s} - V_n\frac{\partial\theta}{\partial r} = -M\left[\frac{\partial f}{\partial\theta} - \nabla\cdot\frac{\partial f}{\partial(\nabla\theta)}\right]. \tag{B.2}$$

For notational convenience, we define $g \coloneqq \frac{\partial f}{\partial\theta}$ and $h\nabla\theta + h'(\boldsymbol{U}\nabla\theta) \coloneqq \frac{\partial f}{\partial(\nabla\theta)}$. The scalar functions $h = h(\nabla\theta, |\nabla\theta|)$ and $h' = h'(\nabla\theta, |\nabla\theta|)$ are obtained from Eq. (3.4) as:

$$h = \frac{B_{aniso}}{|\theta^+ - \theta^-|^2}(2w + w'|\nabla\theta|^2) + \frac{B_{iso}\alpha}{|\theta^+ - \theta^-|^2}[2(1-w) - w'|\nabla\theta|^2] - B_{iso}\beta w'c, \tag{B.3}$$

$$h' = \frac{B'_{aniso}}{|\theta^+ - \theta^-|^2}w|\nabla\theta|^2, \tag{B.4}$$

where we define $B'_{aniso}(\boldsymbol{U}\nabla\theta) \coloneqq \frac{\partial B}{\partial(\nabla\theta)}$ and $w'\nabla\theta \coloneqq \frac{\partial w}{\partial(\nabla\theta)}$. The divergence term $\nabla\cdot\frac{\partial f}{\partial(\nabla\theta)}$ in the curvilinear coordinates $(r,s)$ is given by $\nabla\cdot\frac{\partial f}{\partial(\nabla\theta)} = \nabla\cdot(h\nabla\theta) + \nabla\cdot[h'(\boldsymbol{U}\nabla\theta)]$, with,

$$\nabla\cdot(h\nabla\theta) = \frac{1}{1+\kappa r}\left[\frac{\partial}{\partial r}\left(h\frac{\partial\theta}{\partial r}\right) + \frac{\kappa}{1+\kappa r}h\frac{\partial\theta}{\partial r}\right] + \frac{1}{1+\kappa r}\frac{\partial}{\partial s}\left(\frac{1}{1+\kappa r}h\frac{\partial\theta}{\partial s}\right), \tag{B.5}$$

and

$$\nabla\cdot[h'(\boldsymbol{U}\nabla\theta)] = -\frac{1}{1+\kappa r}\frac{\partial}{\partial r}\left(h'\frac{\partial\theta}{\partial s}\right) + \frac{1}{1+\kappa r}\frac{\partial}{\partial s}\left(h'\frac{\partial\theta}{\partial r}\right). \tag{B.6}$$

Here, $\kappa$ is the interface curvature, which is independent of $\eta$ in the GB region. Since $r$ is small in the GB region, we use the Taylor expansion:

$$\frac{1}{1+\kappa r} = 1 - \kappa r + O(r^2). \tag{B.7}$$

Substituting Eq. (B.7) into Eqs. (B.5) and (B.6), we obtain:

$$\nabla\cdot(h\nabla\theta) \approx \frac{\partial}{\partial r}\left(h\frac{\partial\theta}{\partial r}\right) + \kappa h\frac{\partial\theta}{\partial r} + \frac{\partial}{\partial s}\left(h\frac{\partial\theta}{\partial s}\right) + O(r), \tag{B.8}$$

$$\nabla\cdot[h'(\boldsymbol{U}\nabla\theta)] \approx -\frac{\partial}{\partial r}\left(h'\frac{\partial\theta}{\partial s}\right) + \frac{\partial}{\partial s}\left(h'\frac{\partial\theta}{\partial r}\right) + O(r). \tag{B.9}$$

We expand the orientation field in the GB region as a power series of a small parameter $\varepsilon$:

$$\theta = \theta_0 + \varepsilon\theta_1 + O(\varepsilon^2), \tag{B.10}$$

which yields:

$$\nabla\theta = \nabla\theta_0 + \varepsilon\nabla\theta_1 + O(\varepsilon^2), \tag{B.11}$$

$$|\nabla\theta| = |\nabla\theta_0 + \varepsilon\nabla\theta_1 + O(\varepsilon^2)| \approx |\nabla\theta_0| + \varepsilon\frac{\nabla\theta_0\cdot\nabla\theta_1}{|\nabla\theta_0|} + O(\varepsilon^2). \tag{B.12}$$

Accordingly, the functions $g$, $h$, and $h'$ can be expanded as:

$$g = g_0 + \varepsilon g_1 + O(\varepsilon^2), \tag{B.13}$$

$$h(\theta, |\nabla\theta|) = h_0 + \varepsilon h_1 + O(\varepsilon^2), \tag{B.14}$$

$$h'(\theta, |\nabla\theta|) = h'_0 + \varepsilon h'_1 + O(\varepsilon^2), \tag{B.15}$$

where $h_0 = h(\theta_0, |\nabla\theta_0|)$ and $h_1 = h_\theta(\theta_0, |\nabla\theta_0|)\theta_1 + h_{|\nabla\theta|}(\theta_0, |\nabla\theta_0|)\frac{\nabla\theta_0\cdot\nabla\theta_1}{|\nabla\theta_0|}$, with analogous expressions for $h'_0$ and $h'_1$. All zeroth order terms are independent of tangential variations and are functions of $r$ only. In the GB region, the length scale in the tangential direction $s$ is much larger than that in the normal direction $r$. We therefore introduce the scaled coordinates:

$$\eta = \frac{r}{\varepsilon}, \tag{B.16}$$

while $s$ remains unchanged. With this scaling, $\eta \in (-\infty, +\infty)$ across the GB region, $\frac{\partial}{\partial r}\sim O(\varepsilon^{-1})$, and $\frac{\partial}{\partial s}\sim O(1)$. The relevant terms in Eq. (B.2) become:

$$\frac{\partial\theta}{\partial r} = \frac{1}{\varepsilon}\frac{\partial\theta}{\partial\eta} = \frac{1}{\varepsilon}\frac{\partial\theta_0}{\partial\eta} + \frac{\partial\theta_1}{\partial\eta} + O(\varepsilon), \qquad \frac{\partial\theta}{\partial s} = O(\varepsilon), \tag{B.17}$$

$$\frac{\partial}{\partial r}\left(h\frac{\partial\theta}{\partial r}\right) = \frac{1}{\varepsilon^2}\frac{\partial}{\partial\eta}\left(h\frac{\partial\theta}{\partial\eta}\right) = \frac{1}{\varepsilon^2}\frac{\partial}{\partial\eta}\left(h_0\frac{\partial\theta_0}{\partial\eta}\right) + \frac{1}{\varepsilon}\frac{\partial}{\partial\eta}\left(h_0\frac{\partial\theta_1}{\partial\eta} + h_1\frac{\partial\theta_0}{\partial\eta}\right) + O(1), \tag{B.18}$$

$$\kappa h\frac{\partial\theta}{\partial r} = \frac{1}{\varepsilon}\kappa h_0\frac{\partial\theta_0}{\partial\eta} + O(1), \tag{B.19}$$

$$\frac{\partial}{\partial s}\left(h\frac{\partial\theta}{\partial s}\right) = O(\varepsilon), \tag{B.20}$$

$$\frac{\partial}{\partial r}\left(h'\frac{\partial\theta}{\partial s}\right) = O(1), \qquad \frac{\partial}{\partial s}\left(h'\frac{\partial\theta}{\partial r}\right) = O(1), \tag{B.21}$$

$$\left.\frac{\partial\theta}{\partial t}\right|_{r,s} = O(1). \tag{B.22}$$

Substituting Eqs. (B.17)-(B.22) into Eq. (B.2) yields:

$$-V_n \frac{1}{\varepsilon}\frac{\partial\theta_0}{\partial\eta} = -M\left[-\frac{1}{\varepsilon^2}\frac{\partial}{\partial\eta}\left(h_0\frac{\partial\theta_0}{\partial\eta}\right) - \frac{1}{\varepsilon}\frac{\partial}{\partial\eta}\left(h_0\frac{\partial\theta_1}{\partial\eta} + h_1\frac{\partial\theta_0}{\partial\eta}\right) - \frac{1}{\varepsilon}\kappa h_0\frac{\partial\theta_0}{\partial\eta}\right], \tag{B.23}$$

where only the leading and next order terms are kept. Rearranging, we obtain at leading order $O(\varepsilon^{-2})$:

$$\frac{\partial}{\partial\eta}\left(h_0\frac{\partial\theta_0}{\partial\eta}\right) = 0. \tag{B.24}$$

Integrating both sides yields:

$$\frac{\partial\theta_0}{\partial\eta} = \frac{A}{h_0}, \tag{B.25}$$

where $A$ is the integration constant. In the bulk limit, $\theta^+ = \theta^- = \lim_{\eta\to\infty}\theta_0$ , $\lim_{\eta\to\infty}\frac{\partial\theta_0}{\partial\eta} = 0$ , and $\lim_{\eta\to\infty}\frac{1}{h_0} = 0$ . Therefore, the constant $A$ only needs to be bounded for the leading-order equation (B.24) to be satisfied over the entire domain. At next order $O(\varepsilon^{-1})$, we obtain:

$$M\frac{\partial}{\partial\eta}\left(h_0\frac{\partial\theta_1}{\partial\eta} + h_1\frac{\partial\theta_0}{\partial\eta}\right) = -(V_n + M\kappa h_0)\frac{\partial\theta_0}{\partial\eta}. \tag{B.26}$$

Integrating Eq. (B.26) over the entire domain yields:

$$\left.\left(h_0\frac{\partial\theta_1}{\partial\eta} + h_1\frac{\partial\theta_0}{\partial\eta}\right)\right|_{-\infty}^{+\infty} = -\int_{-\infty}^{+\infty}\left(\frac{1}{M}V_n + \kappa h_0\right)\frac{\partial\theta_0}{\partial\eta}\mathrm{d}\eta. \tag{B.27}$$

In the bulk limit, $\lim_{\eta\to\infty}\frac{\partial\theta_1}{\partial\eta} = \lim_{\eta\to\infty}\frac{\partial\theta_0}{\partial\eta} = 0$. Therefore, the left side of Eq. (B.27) vanishes. For the right side of Eq. (B.27), we first note that:

$$\int_{-\infty}^{+\infty}\frac{\partial\theta_0}{\partial\eta}\mathrm{d}\eta = \theta^+ - \theta^- = \Delta\theta. \tag{B.28}$$

Moreover, in Section III A, the orientation mobility $M$ is chosen to be constant, allowing $\frac{1}{M}$ to be taken outside the integral. Thus, Eq. (B.27) reduces to:

$$\frac{1}{M}V_n\Delta\theta + \kappa\int_{-\infty}^{+\infty}h_0\frac{\partial\theta_0}{\partial\eta}\mathrm{d}\eta = 0. \tag{B.29}$$

Finally, using the relationship $V_n = -M_{GB}\gamma\kappa$, the GB mobility is given by:

$$M_{GB} = M\frac{\int_{-\infty}^{+\infty}h_0\frac{\partial\theta_0}{\partial\eta}\mathrm{d}\eta}{\gamma\Delta\theta}, \tag{B.30}$$

where $\gamma$ is the GB energy density. Since the integrand is localized to the GB region and has been shown to be bounded from the leading-order equation, the GB mobility defined in Eq. (B.30) is finite.

## APPENDIX C

For $m$-fold symmetry, the classical anisotropic GB energy is given by:

$$\gamma_{gb} = \gamma_{gb}^0[1 + E_m\cos m(\phi - \phi_0)], \tag{C.1}$$

where $E_m$ is the anisotropic coefficient for $\gamma_{gb}$ , $m$ represents $m$-fold symmetry of both $\gamma_{gb}$ and the constructed Wulff shape, and $\gamma_{gb}^0$ and $\phi_0$ are fitting parameters. The variable $\phi \in [0,2\pi]$ represents the angular coordinate of $\gamma_{gb}$. The Wulff shape can be constructed by the following steps:

1) Plot the function $\gamma_{gb}$ in polar coordinates $(p,\phi)$, where $p(\phi) = \gamma_{gb}^0[1 + E_m\cos m(\phi - \phi_0)]$;
2) For each point $P$ on the polar plot of $\gamma_{gb}$ , construct a line through $P$ that is normal to the line emanating from the origin to $P$;
3) Construct the inner convex envelope of all such lines.

Next, we determine the Cartesian coordinate of the envelope. Let $T = (r,\Phi)$ be a point on the GB and let $T = (x(\phi), y(\phi))$ be the corresponding Cartesian coordinates. By definition, the line through $P$ on the GB energy plot must be tangent to the GB at $T$. This follows the geometrical relation shown in **FIG C1**:

$$p = r\cos(\phi - \Phi) = x\cos\phi + y\sin\phi. \tag{C.2}$$

Noting that the GB energy normal $[\cos\phi \quad \sin\phi]^T$ is orthogonal to the GB tangent $[x_\phi \quad y_\phi]^T$, we take the derivative to $p$ with respect to $\phi$:

$$p_\phi = -x\sin\phi + y\cos\phi. \tag{C.3}$$

Combining Eqs. (C.2) and (C.3), we obtain expressions for $x$ and $y$:

$$x = p\cos\phi - p_\phi\sin\phi, \qquad y = p\sin\phi + p_\phi\cos\phi, \tag{C.4}$$

where $p = \gamma_{gb}^0[1 + E_m\cos m(\phi - \phi_0)]$ and $p_\phi = -\gamma_{gb}^0 m E_m\sin m(\phi - \phi_0)$.

The curvature of a Wulff construction is given by:

$$\kappa_W(\phi) = \frac{x_\phi y_{\phi\phi} - y_\phi x_{\phi\phi}}{\left(x_\phi^2 + y_\phi^2\right)^{\frac{3}{2}}}. \quad (C.5)$$

By substituting from Eq. (C.4), the corresponding radius of curvature becomes:

$$\rho(\phi) = |-1 + E_m(m^2 - 1)\cos m(\phi - \phi_0)|. \quad (C.6)$$

For the construction to be convex for all orientations, the radius of curvature must not change sign over $\phi$. This requires:

$$-1 + E_m(m^2 - 1)\cos m(\phi - \phi_0) \leq 0 \text{ or } \geq 0, \quad \forall\phi \in [0,2\pi), \quad (C.7)$$

which leads to the condition:

$$E_m \leq \frac{1}{m^2 - 1}. \quad (C.8)$$

For example, in the case of four-fold symmetry ($m = 4$), the convexity condition becomes $E_4 \leq \frac{1}{15}$.

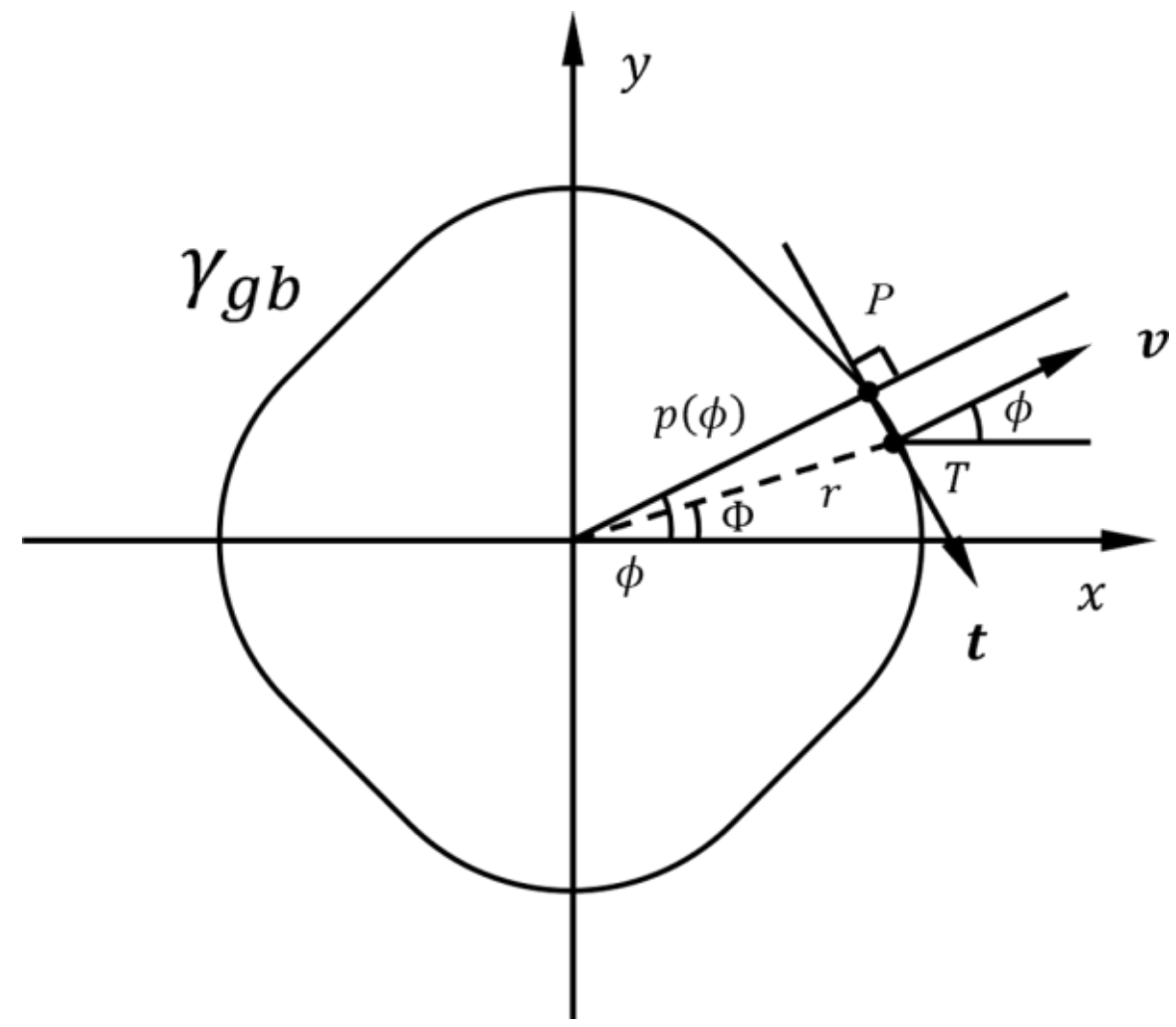

**FIG C1.** A schematic diagram for the analytical Wulff construction on the GB energy profile.